 \documentclass[aps, twocolumn, showpacs, amsmath]{revtex4}
\usepackage{dcolumn}
\usepackage{bm}
\usepackage{graphicx}
\usepackage{color}
\usepackage{subfigure}
\usepackage{hyperref}
\usepackage{latexsym}
\usepackage{amsthm}
\usepackage{amssymb}
\usepackage{braket}
\DeclareGraphicsExtensions{.jpg,.pdf, .mps, .png, .eps, .ps, .EPS,.gif}

\begin{document}
\def\be{\begin{equation}}
\def\ee{\end{equation}}

\def\bc{\begin{center}}
\def\ec{\end{center}}
\def\bea{\begin{eqnarray}}
\def\eea{\end{eqnarray}}
\newcommand{\avg}[1]{\langle{#1}\rangle}
\newcommand{\Avg}[1]{\left\langle{#1}\right\rangle}

\def\ie{\textit{i.\,e.,}}
\def\etal{\textit{et al.}}
\def\m{\vec{m}}
\def\G{\mathcal{G}}

\newcommand{\davide}[1]{{\bf #1}}
\newcommand{\gin}[1]{{\bf\color{green}#1}}
\newcommand{\bob}[1]{{\bf\color{red}#1}}

\title{ Renormalization group for link percolation on planar hyperbolic manifolds}

\author{Ivan Kryven}
\affiliation{Mathematical Institute, Utrecht University, PO Box 80010, 3508 TA Utrecht,  the Netherlands}
\author{Robert M. Ziff}
\affiliation{Center for the Study of Complex Systems and Department of Chemical Engineering, University of Michigan, Ann Arbor, Michigan 48109-2136, USA}
\author{Ginestra Bianconi}
\affiliation{School of Mathematical Sciences, Queen Mary University of London, London, E1 4NS, United Kingdom\\The Alan Turing Institute, 96 Euston Rd, London NW1 2DB, United Kingdom}
\begin{abstract}
Network geometry is currently a topic of growing scientific interest, as it opens the possibility to explore and interpret the interplay between structure and dynamics of complex networks using geometrical arguments. However, the field is still in its infancy.  In this work we investigate the role of network geometry in determining the nature of the percolation transition in planar hyperbolic manifolds. In Ref.\ \cite{hyperbolic_Ziff},  S. Boettcher, V. Singh, R. M. Ziff    have shown that a special type of two-dimensional hyperbolic manifolds, the Farey graphs, display a discontinuous transition for ordinary link  percolation. Here we investigate using  the renormalization group the critical properties of link percolation on a wider class of two-dimensional hyperbolic deterministic and random manifolds constituting the skeletons of two-dimensional cell complexes. These hyperbolic manifolds  are built iteratively by subsequently gluing $m$-polygons to single edges.   We show that when the size $m$ of the polygons is drawn from a  distribution $q_m$ with asymptotic power-law  scaling $q_m\simeq Cm^{-\gamma}$ for $m\gg1$, different universality classes can be observed depending on the value of the power-law exponent $\gamma$. Interestingly,  the percolation transition is hybrid for $\gamma\in (3,4)$ and  becomes continuous for $\gamma \in (2,3]$.
\end{abstract}


\maketitle

\section{Introduction}

The field of network topology and geometry is gaining increasing attention \cite{Perspectives,Lambiotte}.
When modelling network geometry, higher-order networks such as simplicial and cell complexes are a natural choice \cite{NGF,Polytopes,Emergent,Aste,Farey,apollonian1}. Simplicial complexes are formed by simplices that describe the interaction between  one, two or more than two nodes and they include nodes, links, triangles, and tetrahedra. Cell complexes are also formed by high-dimensional building blocks but these building blocks can be any polytopes and do not need to be formed by a set of fully connected nodes.
Models of  simplicial and cell complexes   growing by the subsequent addition of simplices and polytopes include Farey graphs \cite{Farey}, Apollonian networks \cite{apollonian1} and the recently introduced Network Geometry with Flavor \cite{NGF,Polytopes} which displays   emergent hyperbolic network geometry \cite{Emergent}.
The investigation of the relation between  network geometry of simplicial complexes and dynamics is still in its infancy, and only recently a few works have been  tackling problems in percolation\ \cite{Bianconi_Ziff}, synchronization \cite{Ana,Arenas} and epidemic spreading \cite{Vito}. These results significantly enrich the active debate on the role that hyperbolic networks have on navigability  \cite{Kleinberg,Boguna2, Radicchi}.

The interplay between hyperbolic network geometry and percolation \cite{hyperbolic_Ziff,Bianconi_Ziff,classical,Gu_Ziff,Moore_Mertens,percolation_Apollonian} appears to be very profound. A classical result \cite{classical} of percolation states that in hyperbolic networks as well as in non-amenable graphs percolation exhibits two phase transitions: at the lower percolation threshold an infinite cluster emerges but remains sub-extensive, and at the upper percolation threshold the infinite cluster becomes extensive.
Even more interesting is the  result of  S. Boettcher, V. Singh, R. M. Ziff    \cite{hyperbolic_Ziff}  which shows that on the two-dimensional Farey graph, link (or bond) percolation is discontinuous.
Finally, it has been shown in Ref. \cite{Bianconi_Ziff} that on simplicial complexes of dimension $d$ one can define up to $2d$ topological percolation problems that can display a critical behavior that cannot be predicted by exclusively studying node and link percolation problems on the same network geometries.

Percolation is among the most widely studied critical phenomena on networks \cite{Newman,Havlin_single,Havlin_exponents,Doro_book,Doro_crit,Ziff_review,Kahng_review,Ivan_clusters,kryven2019bond} and for many years it has been argued that percolation could only lead to  second-order phase transitions.
However, in  recent years, there has been an increasing interest in unveiling the basic mechanisms responsible for abrupt percolation transitions in complex networks. In multiplex networks   \cite{Bianconi2018} it has been shown that interdependent percolation leads to hybrid phase transitions \cite{Havlin,Grassberger,Doro_multiplex,Bianconi2018}. In simple networks it has been shown that network processes that are responsible for growing network structure can lead to abrupt but continuous phase transitions (explosive percolation) \cite{dSouza,Ziff_explosive,Doro_explosive,Riordan} as well as truly discontinuous phase transitions \cite{Kahng_choice1,Kahng_choice2}. Moreover, it has been shown  \cite{Ldev}  that in the large deviation theory of percolation, discontinuous phase transitions can be observed if aggravating  configurations of the initial damage are considered. In this context the result of Ref.\  \cite{hyperbolic_Ziff}  is  revealing of the  important interplay between network geometry and the emergence of discontinuous critical behavior of percolation. Interestingly, the discontinuous nature of this link percolation transition is likely to be related to previous work on critical phenomena on one-dimensional systems with long-range interactions \cite{AN,Chayes}. However this relation has so far not been clearly established.

In this work we address the question of the interplay between hyperbolic network geometry and percolation in dimension $d=2$. In particular our goal is to explore how robust is the discontinuity of link percolation observed in Farey graphs with respect to modifications of the geometry of the building blocks of the planar hyperbolic manifold.
Percolation on hierarchical lattices has been widely explored    \cite{Patchy,clusters,Ziff_sierpinski,tricritical,flower_tau} in the literature and advanced renormalization group (RG) techniques \cite{RG,Boettcher_RG,Berker_RG,Boettcher_Ising,Boettcher_Potts} have been developed for studying percolation and critical phenomena on hierarchical networks.  Here we build on this literature to  perform a comprehensive  RG study of link percolation in deterministic and random hyperbolic manifolds in two dimensions. We show that when the size $m$ of the polygons forming the random hyperbolic manifold is drawn from a  distribution $q_m$ with asymptotic power-law  scaling $q_m\simeq Cm^{-\gamma}$ for $m\gg1$,  a rich scenario is observed with the occurrence of different universality classes  depending on the power-law exponent $\gamma$.

The paper is organized as follows: in Sec.\ II we introduce the deterministic and the random hyperbolic manifolds studied in this paper; in Sec.\ III we introduce the main properties of percolation in hyperbolic networks including the upper and lower percolation thresholds and the fractal exponent; in Sec.\ III we determine the equation of the percolation probability and we characterize its critical behavior; in Sec.\ IV we derive the expression for the generating function of the largest component; in Sec.\ V we determine the equations for the fractal exponent and we characterize its critical behavior; in Sec.\ VI we determine the equations for the order parameter of percolation and we establish the universality class of percolation at the upper critical threshold; finally in Sec.\ VII we provide the conclusions.
\section{Deterministic and random hyperbolic manifolds }
\subsection{Deterministic hyperbolic manifolds}

\begin{figure}
\includegraphics[width=\columnwidth]{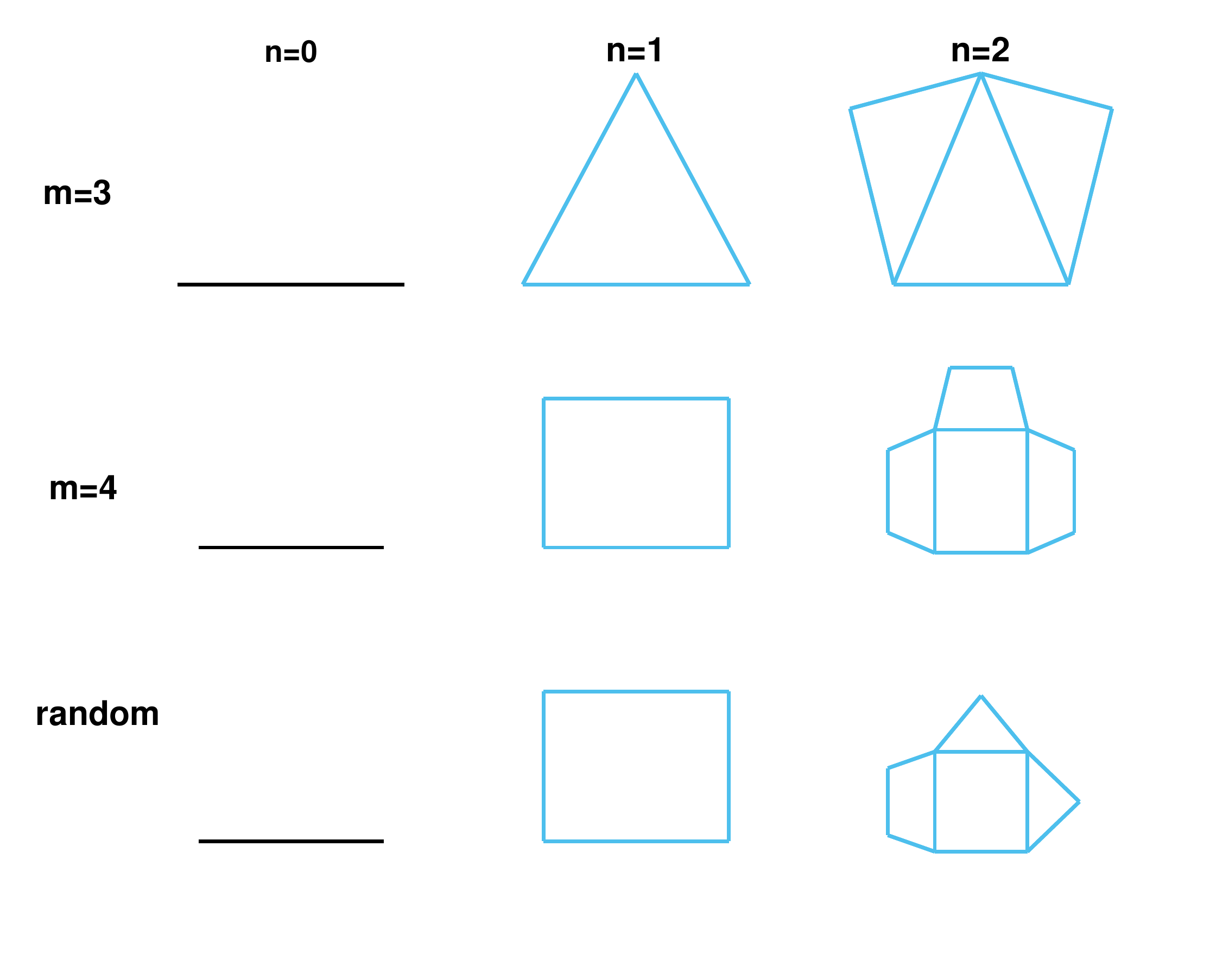}
\caption{The first iterations $n=0,n=1$ and $n=2$ for the constructions of the considered hyperbolic manifolds are here shown in the case in which the hyperbolic manifold is deterministic and only formed by triangles ($m=3$), or squares $(m=4)$ and in the case in which the hyperbolic manifold is random with $q_{m}=0.5\,\delta_{m,3}+0.5\,\delta_{m,4}.$}
\label{fig:0}
\end{figure}

The deterministic  manifolds considered in this paper are 
 infinite hyperbolic simplicial complexes constructed deterministically and iteratively starting from a single link.    At iteration $n=0$ we attach an $m$-polygon (polygon with $m$ links) to the initial link.
At each iteration $n>0$ we attach an $m$-polygon to every link introduced at iteration $n-1$.
At iteration $n$ the number of nodes $N_n$ and links $L_n$ are given by 
\bea
N_n&=&1+(m-1)^n,\nonumber \\
L_n&=&\frac{1}{m-2}\left[(m-1)^{n+1}-1\right].
\eea
The resulting networks constitute the 1-skeleton of $d=2$ hyperbolic cell complexes and describe hyperbolic manifolds. In fact they are 2-connected and every link is incident at most to two polygons.  These hyperbolic manifolds generalize the Farey graphs studied in Ref.\ \cite{hyperbolic_Ziff} which correspond to the case $m=3$, i.\,e., to the case in which the network is constructed by gluing subsequent triangles to links. In Figure $\ref{fig:0}$ we show  the outcome of the first iterations generating the deterministic hyperbolic manifolds with $m=3$ and $m=4$ respectively.

 \subsection{Random hyperbolic manifolds}
The random hyperbolic manifolds considered in this paper are  constructed using an iterative procedure similar to the one used for constructing the deterministic hyperbolic manifolds considered above. However the difference in their construction is that we allow the cell complexes to include polygons of different sizes $m$. To this end we consider an ensemble of random hyperbolic manifolds growing iteratively from a single link.
At iteration $n=0$ we draw a value of $m>2$ from a distribution $q_m$ and we attach an  $m$-polygon to the initial link.
At  iteration $n>0$, for each link added at the previous iteration, we draw a value of $m>2$ from a distribution $q_m$ and we attach an  $m$-polygon to it.

In order to have a well-defined model of random hyperbolic manifolds we will assume here and in the following that the distribution $q_m$ has a well-defined first moment $\avg{m}$.
Under this hypothesis  the expected number of nodes $\bar{N}_n$ and links $\bar{L}_n$ in the manifold at  iteration $n$ are given by 
\bea
\bar{N}_n&=&1+(\avg{m}-1)^n,\nonumber \\
\bar{L}_n&=&\frac{1}{\avg{m}-2}\left[(\avg{m}-1)^{n+1}-1\right].
\label{barN}
\eea
In Figure $\ref{fig:1}$  we show  the outcome of the first iterations generating a random hyperbolic manifolds in which we attach with equal probability either triangles or squares, i.\,e.,  $q_m=0.5\,\delta_{m,3}+0.5\,\delta_{m,4}$ where here and in following $\delta_{a,b}$ indicates the Kronecker delta.
Note that the randomness of these hyperbolic manifolds is only due to the fact that we can add polygons of different number of links $m$. Therefore the random hyperbolic manifolds reduce to the deterministic hyperbolic manifolds defined previously when the distribution $q_{m'}$ is given by $q_{m'}=\delta_{m,m'}$. Consequently  the deterministic hyperbolic manifolds can also be  called  regular hyperbolic manifolds.
\section{Percolation in hyperbolic manifolds}

In link percolation (also called bond percolation) links are removed with probability $f=1-p$ and the fraction $P_{\infty}$ of nodes belonging to the giant component is studied as a function of $p$.
However in the considered hyperbolic manifolds percolation can be also be characterized by  the {\em percolation probability} $T$  and the {\em fractal exponent} $\psi$. The percolation probability $T$  indicates the probability that the two initial nodes present at iteration $n=0$ are connected as $n\to \infty$ {  when links are removed with probability $f=1-p$. The {fractal critical exponent} $\psi$ determines the size of the largest percolation cluster.
In particular the study of $T$ and $\psi$ reveals that link percolation on hyperbolic manifolds \cite{classical} and in general non-amenable graphs  has not just one {but} two percolation thresholds: the lower threshold $p^{\star}$ and  the upper threshold $p_{c}$. 
Therefore, the phase diagram of link percolation on hyperbolic manifolds  includes three regions:
\begin{itemize}
\item
For $p<p^{\star}$ no cluster has infinite size. In this phase the percolation probability is null in the infinite network limit $n\to \infty$ (i.\,e., $T=0$) indicating that the  two initial nodes are not connected.
\item
For $p^{\star}<p<p_c$ the network has a non-zero probability $T>0$ that the initial two nodes are connected.  Moreover, in this phase the  cluster that is connected to the initial two nodes at iteration $n$ has an  expected size $R_n$ which is infinite (increases with the network size $N_n$) but sub-extensive. In other words:  $R_n$ scales with the expected total number of nodes $\bar{N}_n$  in the network as 
\bea
R_n\sim \bar{N}_n^{\psi_n}
\eea
with $0<\psi_n<1$ where the limit 
\bea
\psi=\lim_{n\to \infty}\psi_n
\eea
indicates the {\em fractal critical exponent}.
\item
For $p>p_c$ the network has an extensive cluster. In this case  the probability that the initial two nodes are connected is one, i.\,e., $T=1$, and the fraction $P_{\infty}$ of nodes in the giant component scales like
\bea
P_{\infty} = \lim_{n\to \infty} \frac{R_n}{\bar{N}_n}=O(1),
\eea
or equivalently $\psi=1$.  
\end{itemize}

\section{Percolation probability}

\subsection{Recursive equation for the percolation probability}

Let us indicate with $T_n$  the probability that the  two nodes present at iteration $n=0$  are connected at generation $n$.
Given the iterative process that defines the  hyperbolic manifold, it is easy to show that for the random-$m$ case  $T_n$ satisfies the RG equation
\bea
T_{n+1}=p+(1-p)\sum_{m=3}^{\infty}q_m T_n^{m-1}
\label{Tnr}
\eea
with initial condition $T_0=p$.
  If the link that directly connects the two initial nodes is not damaged, the two  nodes are clearly connected. This event occurs with probability $p$. If the link that directly connects the initial two nodes is instead damaged (events that occur with probability $1-p$), the two nodes are connected only if they are connected by a path that passes through each other node belonging to the polygon added at generation $n=1$. Since the size the polygon added at iteration $n=1$ has size $m$ with probability $q_m$ the latter event  occurs with probability 
$\sum_{m=3}^{\infty} q_m T_n^{m-1}$. 
\begin{figure}[htb]
\includegraphics[width=1\columnwidth]{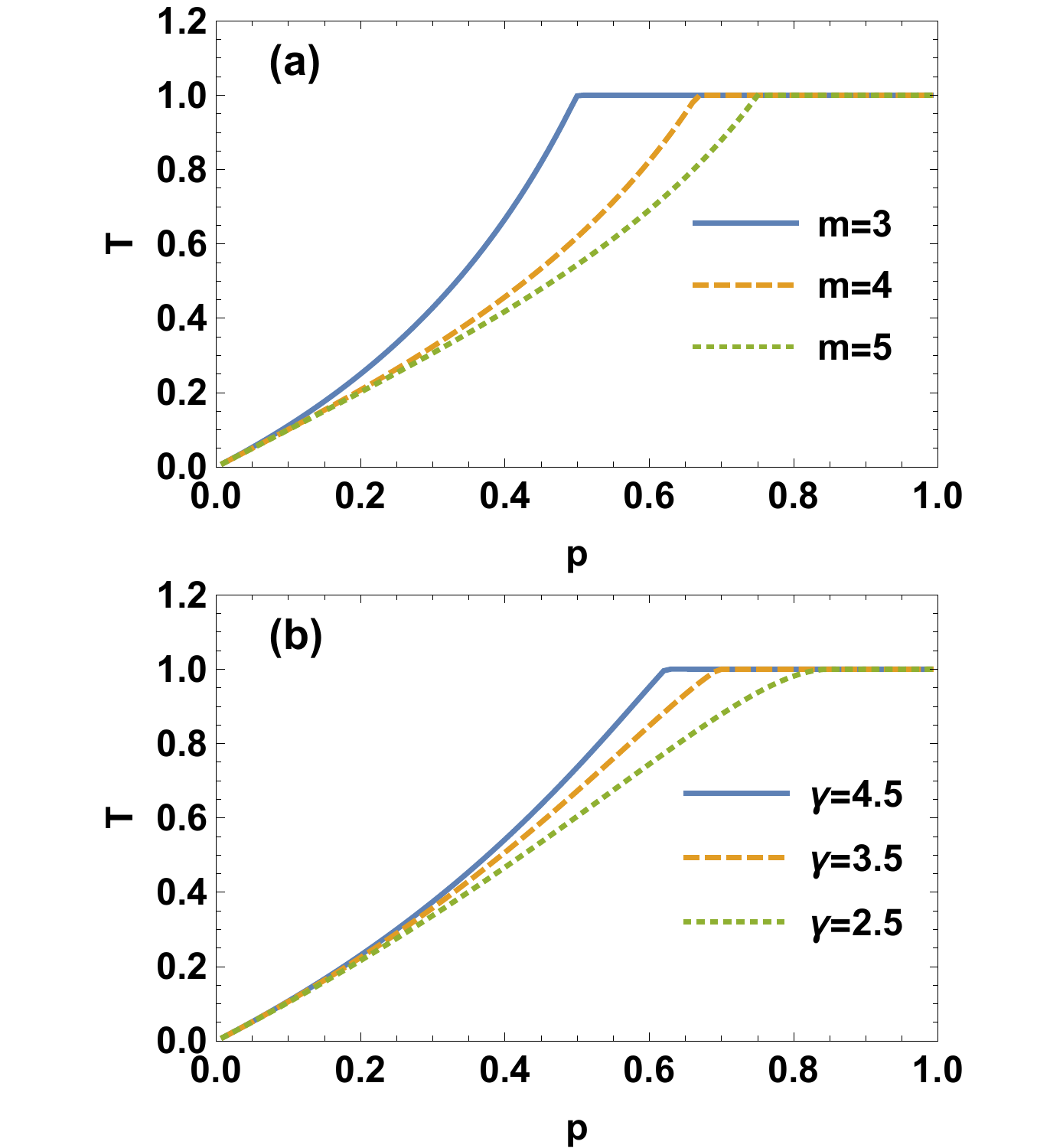}
\caption{The percolation probability $T$ is plotted versus $p$ for deterministic hyperbolic manifolds with $m=3,4,5$ (panel a) and random hyperbolic manifolds with scale-free distribution $q_m=Cm^{-\gamma}$ with  $m\geq 3$ and $\gamma=4.5,3.5,2.5$.}
\label{fig:T}
\end{figure}
For our further calculations  it is useful to express Eq.\ (\ref{Tnr}) as 
\bea
T_{n+1}=F(p,T_n),
\label{RG}
\eea
where 
\bea
F(p,T_n)=p+(1-p)Q(T_n).
\eea
and  where   $Q(T)$ indicates the function
\bea
Q(T)=\sum_{m>2}q_mT^{m-1}.
\label{QT}
\eea

When the number of iterations $n$ diverges $n\to \infty$ we observe that $T_n$ converges to its limiting value $T$, i.\,e., $T_n\to T$, expressing the probability that the initial two nodes are connected in an infinite random hyperbolic manifold. Therefore, starting from the recursive Eq.\ (\ref{Tnr}) we obtain the implicit equation for $T$ given by 
\bea
T=p+(1-p)Q(T)
\label{Tr}
\eea
In Figure $\ref{fig:T}$ we show the linking probability $T$ as a function of $p$ for deterministic hyperbolic manifolds and random hyperbolic manifolds with scale-free distribution $q_m$.
By studying the stability of the solutions $T^{\star}=0$ and $T_c=1$ we observe that Eq.\ ($\ref{Tr}$) identifies the lower percolation threshold $p^{\star}$  and the upper percolation threshold $p_c$ as 
\bea\label{eq:poly:pc}
\begin{array}{clr}
p^{\star}&=0,&  T^{\star}=0;\nonumber \\ 
p_c&=\displaystyle{1-\frac{1}{\avg{m}-1}}, & \  T_c=1.
\end{array}
\eea
For a deterministic hyperbolic manifold with $q_{m'}=\delta_{m,m'}$ we obtain then 
\bea\label{eq:reg:pc}
\begin{array}{clr}
p^{\star}&=0,&  T^{\star}=0;\nonumber \\ 
p_c&=\displaystyle{1-\frac{1}{{m}-1}}, & \  T_c=1.
\end{array}
\eea
It is interesting to compare these results to percolation on a Cayley tree with $m-1$ descendants, which is the dual lattice to the manifolds considered here, as shown in  Figure  \ref{fig:1}.  The probability of a bond presence on a dual lattice is $\tilde{p}= 1 - p$, and indeed, the lower percolation threshold for the Cayley tree is known to be ${1}/(m-1)=1-p_c$, where $p_c$ is given by Eq.~\eqref{eq:reg:pc}. In this case $T=1-S$, where $S$ is the probability that a randomly chosen link in the dual tree leads to the giant component in the direction away from the root.
However, there no known connections between $P_\infty$ and the properties of the dual lattice. 

As a side note, we observe that the existence of the $T=T_c=1$ solution of Eq.\ (\ref{Tr}) is the necessary condition for the presence of an extensive component in the network, i.e. the presence of an upper percolation threshold $p_c$.
Interestingly the presence of this phase at a non-trivial critical point $p=p_c<1$ is not guaranteed in any modification of the hyperbolic manifold construction. In particular if we modify the network construction by
replacing the chain of $m-1$ iterable links by one that has $m-1-k$
iterable links and $k$ non-iterable links we will construct a Kantor set structure. For this structure the linking probability would satisfy
\bea
T=p+(1-p)p^kT^{m-1-k}
\eea
and the $T=T_c=1$ solution would not be allowed unless $p=p_c=1$.

\begin{figure*}
\includegraphics[width=2\columnwidth]{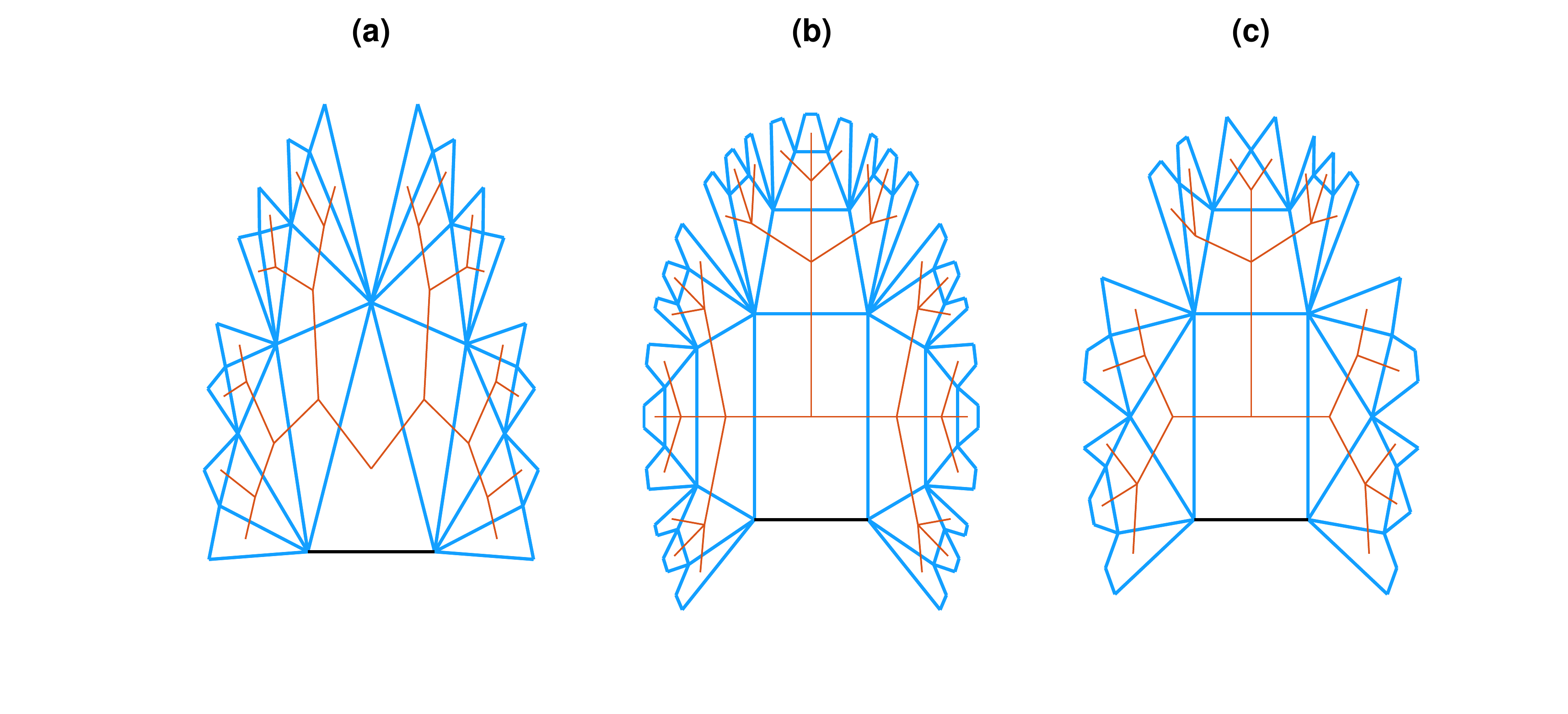}
\caption{(Color online) The results of the first $n$ iterations generating the regular hyperbolic manifolds with $m=3$ (panel a), $m=4$ (panel b) are shown in blue (thick line), together with the results of the first $n$ iteration generating a random hyperbolic manifold with $q_m=1/2\,\delta(m,3)+1/2\,\delta(m,4)$ (panel c).  The initial link is indicated in black and the dual network (the tree) is indicated in red (thin line). The number of iterations $n$ is $n=5$ for panel (a) and $n=4$ for panels (b) and (c).}
\label{fig:1}
\end{figure*}
\subsection{Mapping to percolation of random networks}

In this section we reveal that the  critical behavior of $T$ for $0<p_c-p\ll 1$ can be non-trivial  by considering a mapping between the Eq.\ (\ref{Tr}) for $T$ and percolation on random networks.
 In fact equation (\ref{Tr}) determining the percolation probability $T$  can be mapped to the equation for the probability  $S'$  that a random link reaches a node in the giant component of a random uncorrelated network with degree distribution $P(k)$ and minimum degree $k_{min}\geq 3$.
When nodes are damaged with probability $1-\tilde{p}$ the equation for $S'$ reads 
\bea
S'=\tilde{p}\sum_{k}\frac{k}{\avg{k}}_PP(k)[1-(1-S')^{k-1}],
\label{Sppp}
\eea
where $\avg{k}_P=\sum k P(k)$. 
Therefore at the mathematical level  the equation determining $T$ in random hyperbolic manifolds and the equation determining $S'$ in a random  network can be mapped to each other. 

In this mapping the size of the polygon $m$ corresponds to the degree of a node $k$, the probability distribution $q_m$ corresponds to the distribution of excess degree  $k P(k)/\avg{k}$, $\avg{1/m}=\sum_m q_m/m$ corresponds to $\avg{k}_P$ and $p$ corresponds to $1-\tilde{p}$ (see Table \ref{table_mapping}).

Note that  finite random uncorrelated networks have the structural cutoff, i.e. the maximum degree $k_{max}$  scales like the square root of the number of nodes in the network. However in our mathematical mapping between Eq.\ (\ref{Tr}) and Eq.\ (\ref{Sppp}) there is no prescribed mapping between the number of nodes in the random uncorrelated network and the number of nodes in the hyperbolic manifold, therefore  the mapping described in this section can be performed  for manifolds with an arbitrary large value of  $m_{max}$.

\begin{table}
\begin{tabular}{c c}
Hyperbolic Random Manifold  \qquad & Random Network\\
 $T$					& $1-S'$\\
$p$					&$1-\tilde{p}$\\
 $m$				& $k$\\
 $q_m$				&$\frac{k}{\avg{k}}P(k)$\\
 $\avg{1/m}$ &$\avg{k}_P$
\end{tabular}
\caption{Mathematical mapping between the quantities determining the percolation probability $T$ in the hyperbolic random manifold and the mathematical quantities determining the probability $S'$ that by following a random link of a random  network with degree distribution $P(k)$ we reach a node in the giant component.}
\label{table_mapping}
\end{table}

In this mapping the upper percolation threshold $p_c$ maps to the well-known percolation threshold $\tilde{p}_c$ for the random network. In fact we have  that $p_c$ is determined by the equation
\bea
(1-p_c)\left[\sum_{m=3}^{\infty}q_m(m-1)\right]=1
\eea
By substituting 
\bea
1-p_c & \to &\tilde{p}_c\nonumber \\
m& \to& k\nonumber \\
q_m & \to & \frac{k}{\avg{k}_P}P(k).
\eea
we obtain 
\bea
\tilde{p}_c\left[\sum_{k>2} \frac{k(k-1)}{\avg{k}}P(k)\right]=1.
\eea
{or
\bea
\tilde{p}_c = \frac{\avg{k}_P}{\avg{k^2}_P - \avg{k}_P}
\eea
where the averages $\avg{\dots}_P$  are over the distribution $P(k)$.}
It is well-known that the percolation threshold and the critical behavior of percolation are strongly affected by power-law degree distributions for random networks, which display anomalous critical exponents \cite{Newman,Havlin,Havlin_exponents,Doro_crit,Doro_book}.  
For example, for the distribution $q_m\simeq Cm^{-\gamma}$ for $m\gg1 $ and $\gamma\to 2^{+}$, the percolation threshold $p_c\to 1$. Therefore in this limit the network displays an extensive giant component only when the fraction of damaged nodes $f=1-p=0$. 
This regime corresponds to the regime in which random networks have a degree distribution decaying as $P(k)\simeq Ck^{-\tilde{\gamma}}$ with $\tilde{\gamma}\to 3^{+}$ and percolation threshold $\tilde{p}_c\to 0$. 
\subsection{Critical behavior of the linking probability}
Using the mapping between random hyperbolic manifolds and random networks we can  also  determine the different critical behaviors for $T$ in the deterministic and in the random hyperbolic manifolds.
To this end we will consider different scenarios.

First, we consider the case of a generic distribution $q_m$ displaying finite average moments $\avg{m}$ and $\avg{m^2}$. This case clearly includes the deterministic hyperbolic manifold.
For $0<p_c-p\ll 1$ the function $Q(T)$ can be expanded in powers of $\Delta T=T-T_c$ as we have $|\Delta T|\ll1$, yielding
\bea
Q(T)&=& 1+(\avg{m}-1)(\Delta T)+\frac{1}{2}\avg{(m-1)(m-2)}(\Delta T)^2\nonumber \\&&+{o}((\Delta T)^2).
\eea
Using this expansion in 
Eq.\ (\ref{Tr}), we get
\bea\label{eq:polyQ:expansion}
T=1-A\left(p_c-p\right)^{\beta}+ o(\left(p_c-p\right)^{\beta}),
\label{ATp}
\eea
where the   critical exponent $\beta$ is given by 
\bea
\beta=1,
\eea
and where $A$ is a constant given by 
\bea
A=\frac{2 (\Avg{m} -1)^2}{\Avg{(m-2)(m-1)}}.
\label{Adef}
\eea
In particular in the case of a deterministic hyperbolic manifold $A=A_m$ where $A_m$ is given by 
\bea
A_m=2\frac{m -1}{ m-2}.
\eea

Secondly we consider the case of distributions $q_m$ with power-law asymptotic behavior 
\bea
q_m\simeq Cm^{-\gamma},
\eea
for $m\gg1 $ where the range of possible exponents is $\gamma\in (2,\infty]$ as we need to guarantee that $\avg{m}$ is finite as mentioned above.     
Special attention is devoted in particular to 
 power-law exponents $\gamma\leq 3$ corresponding to a diverging moment $\avg{m^2}$. { In particular in this regime, by following techniques already developed for random networks  \cite{Newman,Havlin,Havlin_exponents,Doro_crit,Doro_book},  we found a series of anomalous critical exponents.} 
\begin{itemize}
\item[(a)]{\it Case $\gamma>3$}\\
For $\gamma>3$ both moments $\avg{m}$ and $\avg{m^2}$ converges so this case follow in the universality class of the case we have studied previously. Specifically in this case we can expand $Q(T)$ for $|\Delta T|\ll1$ getting 
 \bea
 Q(T)&=& 1+(\avg{m}-1)\Delta T+a_{\gamma}(\Delta T)^2\nonumber \\&&+o((\Delta T)^2)).
 \eea
with $a_{\gamma}=\avg{(m-1)(m-2}/2$ (see Table $\ref{table:acd}$ for its expression in the case of a pure power-law distribution $q_m$).
By inserting this asymptotic expansion in Eq.\ (\ref{Tr}) we obtain
\bea
T= 1- A_{\gamma}\left({p_c-p}\right)^{\beta}+o(({p_c-p})^{\beta}),
\label{Tmf}
\eea
with $A_{\gamma}=\avg{m-1}/a_{\gamma}$ and 
\bea
\beta=1.
\eea

 \item[(b)]{\it Case $\gamma=3$}\\
 For $\gamma=3$ we can expand $Q(T)$ for $|\Delta T|\ll1$ getting 
 \bea
 Q(T)&=& 1+\avg{m-1}\Delta T+a_{\gamma}(\Delta T)^2\ln|\Delta T|\nonumber \\&&{\mathcal O}((\Delta T)^2).
 \eea
with $a_{\gamma}$ indicating a constant (see Table $\ref{table:acd}$ for its expression in the case of a pure power-law distribution $q_m$).
By inserting this asymptotic expansion in Eq.\ (\ref{Tr}) we obtain
\bea
T= 1- A_{\gamma}\left({p_c-p}\right)\left[-\ln\left({p_c-p}\right)\right]^{-1}+{\mathcal O}({p_c-p}),
\label{T3}
\eea
with $A_{\gamma}=-(\avg{m}-1)^2/a_{\gamma}$.
\item[(c)]{\it Case $\gamma\in (2,3)$}\\
 For $\gamma\in (2,3)$ we can expand $Q(T)$ for $|\Delta T|\ll1$ getting 
 \bea
 Q(T)&=& 1+(\avg{m}-1)\Delta T+a_{\gamma}|\Delta T|^{\gamma-1}\nonumber \\&&+o(|\Delta T|^{\gamma-1}).
 \eea
with $a_{\gamma}$ indicating a constant (see Table $\ref{table:acd}$ for its expression in the case of a pure power-law distribution $q_m$).
By inserting this asymptotic expansion in Eq.\ (\ref{Tr}) we obtain
\bea
T=1- A_{\gamma}\left({p_c-p}\right)^{\beta}+ {o}(({p_c-p})^{\beta}),
\label{T23}
\eea
with $A_{\gamma}=[(\avg{m}-1)^2/a_{\gamma}]^{\beta}$ and
\bea
\beta=\frac{1}{\gamma-2}.
\eea
\end{itemize}
In Table \ref{table:expansions} we summarize the different scaling behaviors observed for different values of $\gamma$.

\begin{table}
\bgroup
\def\arraystretch{1.5}
\begin{center}
\begin{tabular}{ccc}
$\gamma$\hspace{0.5cm} 		&   Expansion of $T_c-T$& $\beta$\hspace{0.5cm}		 \\
\hline
$\gamma>3$	&   	$A_{\gamma}\left(p_c-p\right)$						&$1$	\\
$\gamma=3$		&   	$A_{\gamma}\left({p_c-p}\right)\left[-\ln\left({p_c-p}\right)\right]^{-1}$		& N/A	\\
$2<\gamma<3$	&  	$A_{\gamma} (p_c-p)^{\frac{1}{\gamma-2}}$				&$\frac{1}{\gamma-2}$		
\end{tabular}\\
\end{center}
\caption{
 Expansions of $T_c-T$  for $p=p_c-\epsilon$ and $0<\epsilon\ll1 $. Here, $\gamma>2$ denotes the exponent of the asymptotic power-law scaling of the $q_m$ distribution, i.\,e.,  $q_m\simeq C m^{-\gamma}$ for $m\gg 1$. For the case $\gamma=3$ the dynamical critical exponent $\beta$ is not defined (N/A) since the scaling of $T_c-T$ is linear  with logarithmic corrections.}
\label{table:expansions}
\egroup
\end{table}%

\section{Generating functions for finite components}
\subsection{General framework}

In order to fully characterize the percolation transition in the considered hyperbolic manifolds we follow the theoretical approach proposed by Boettcher, Singh and Ziff in Ref.\ \cite{hyperbolic_Ziff} and we investigate the properties of the  generating functions  $\hat{T}_n(x)$ and $\hat{S}_n(x,y)$.
In the  hyperbolic manifolds obtained at  iteration $n$ the  function $\hat{T}_n(x)$ is the generating function of the  
number of nodes in the connected component linked to both initial nodes. The function $\hat{S}_n(x,y)$  is the generating function for the sizes of the two connected components linked  exclusively to one of the two initial nodes. 
These generating functions can be expressed as 
\bea
\hat{T}_n(x)&=&\sum_{\ell=0}^{\infty}t_n(\ell) x^{\ell},\nonumber \\
\hat{S}_n(x,y)&=&\sum_{\ell,\bar{\ell}}s_n(\ell,\bar{\ell})x^{\ell}y^{\bar{\ell}},
\eea
where  $t_n(\ell)$ indicates  the distribution of the number of nodes $\ell$ connected to the two initial nodes and  $s_n(\ell,\bar{\ell})$  indicates  the joint distribution of  the number of nodes $\ell$ connected exclusively to a given initial node and the number of nodes $\bar{\ell}$ connected exclusively to the other initial node.

The size $R_n$ of the connected component linked to the initial  two nodes at iteration $n$ is given by 
\bea
R_n=\left.\frac{d\hat{T}_n(x)}{dx}\right|_{x=1}.
\eea
By explicitly deriving $R_n$ it can be shown that    for $n\gg1 $, $R_n$ scales like
\bea
R_n\sim \bar{N}_n^{\psi},
\eea
with $\psi\in (0,1]$ for $p>p^{\star}$.
In the section below we will consider the recursive equation that can be used to determine the generating functions  $\hat{T}_n(x)$ and $\hat{S}_n(x,y)$ for the deterministic and the random hyperbolic manifolds while the next section will be devoted to the evaluation of the fractal exponent $\psi$.

\subsection{Deterministic hyperbolic manifold}
Here we establish the equations determining the generating functions $\hat{T}_n(x)$ and $\hat{S}_n(x,y)$ for   the deterministic hyperbolic manifold formed by gluing $m$-polygons together.
These  recursive equations for $\hat{T}_n(x)$ and $\hat{S}_n(x,y)$ start from the initial condition $T_{0}(x)=1-\hat{S}_{0}(x,y) =p$  and read 
\begin{widetext}
\bea
\hat{T}_{n+1}(x)&=&p\left[x^{m-2}\hat{T}_n^{m-1}(x)+(m-1)x^{m-2}\hat{T}_n^{m-2}(x)\hat{S}_n(x,x)+ \sum _{i=0}^{m-3}(i+1)x^{i} \hat{T}_n^{i}(x) \hat{S}_n^2(x,1)\right]+(1-p) x^{m-2} \hat{T}_n^{m-1}(x),\nonumber \\
\hat{S}_{n+1}(x,y)&=&(1-p) \left[  \sum _{i=0}^{m-2} x^i y^{m-2-i} \hat{T}^i_n (x)\hat{T}^{m-2-i}_n(y)\hat{S}_n(x,y) +\sum_{ i=0}^{m-3}\sum_{j=0}^{m-3-i} x^i y^j \hat{T}^i_n(x) \hat{T}^j_n(y)\hat{S}_n(x,1) \hat{S}(1,y)\right].
\label{gen_m}
\eea
\end{widetext}
The generating function $\hat{T}_{n+1}(x)$ of the size of the connected component joining  the two initial nodes  should consider only contributions from configurations in which the link that connects the two initial nodes  is not damaged, or the cases in which the initial  link is damaged but the two initial nodes are  connected through  paths  that pass through each other node of the $m$-polygon added at iteration $n=1$. The generating function $\hat{S}_{n+1}(x,y)$ should instead take into account only contributions from configurations in which the initial link between the two initial nodes is damaged and there exist no alternative path connecting the two initial nodes. 

In order to derive these equations it is possible to consider the simple cases in which $m=3$ and $m=4$ and by induction  prove the general formula for a generic value of $m$. The recursive equations for $m=3$ and 4 can be easily derived diagrammatically using the diagram for $\hat{T}_n(x)$ and $\hat{S}_n(x,y)$ shown in Figure $\ref{fig:diagrams1}$.
For the case $m=3$ these equations reduce to the equations derived in Ref.\ \cite{hyperbolic_Ziff} for the Farey graph, which read
\begin{widetext}
\bea
\hat{T}_{n+1}(x)&=&p\left\{x\hat{T}_n^2(x)+2x\hat{T}_n(x)\hat{S}_n(x,x)+\hat{S}_n(1,x)\hat{S}_n(1,x)\right\}+(1-p)x\hat{T}_n^2(x)\nonumber \\
\hat{S}_{n+1}(x,y)&=&(1-p)\left\{x\hat{T}_n(x)\hat{S}_n(x,y)+y\hat{S}_n(x,y)\hat{T}_n(y)+\hat{S}_n(x,1)\hat{S}_n(y,1)\right\}.
\eea
\end{widetext}
In the case $m=4$ instead the equations can be derived by using a diagramatic representation of the configurations that contribute to $\hat{T}_{n+1}(x)$ and $\hat{S}_{n+1}(x,y)$  (see Figures \ref{fig:diagrams2} and  \ref{fig:diagrams3})  obtaining
\begin{widetext}
\bea
\hat{T}_{n+1}(x)&=&p[x^2\hat{T}_n^3(x)+3x^2\hat{T}_n^2(x)\hat{S}_n(x,x)+2x\hat{T}_n(x)\hat{S}_n(x,1)\hat{S}_n(x,1)+\hat{S}_n(x,1)\hat{S}_n(x,1)]+(1-p)x^2\hat{T}_n^3(x)\nonumber \\
\hat{S}_{n+1}(x,y)&=&(1-p)\left[x^2\hat{T}_n^2(x)\hat{S}_n(x,y)+y^2\hat{T}_n^2(y)\hat{S}_n(x,y)+xy\hat{T}_n(x)\hat{S}_n(x,y)\hat{T}_n(y)+x\hat{T}_n(x)\hat{S}_n(x,1)\hat{S}_n(y,1)\right.\nonumber \\
&&\left.+y\hat{T}_n(y)\hat{S}_n(y,1)\hat{S}_n(x,1)+\hat{S}_n(x,1)\hat{S}_n(y,1)\right]
\eea
\end{widetext}

\subsection{Random  hyperbolic manifolds}
The recursive equations for the generating functions $\hat{T}_n(x)$  and $\hat{S}_n(x,y)$ of the random hyperbolic  manifold can be easily derived from the Eqs.\ (\ref{gen_m}) for the same generating function in a deterministic $m$-polygon hyperbolic manifolds by averaging over the distribution $q_m$. In this way we obtain
\begin{widetext}
\bea
\hat{T}_{n+1}(x)&=&\sum\limits_{m=3}^\infty q_m \left( x^{m-2} \hat{T}_n^{m-1}(x)+p (m-1)x^{m-2} \hat{T}_n^{m-2}(x) \hat{S}_n(x,x)  +p  \sum_{i=0}^{m-3} (i+1)x^{i} \hat{T}_n^{i}(x)\hat{S}_n^2(x,1)\right),\\
\hat{S}_{n+1}(x,y)&=&(1-p) \sum\limits_{m=3}^\infty q_m\left( \sum _{i=0}^{m-2} x^i y^{m-2-i} \hat{T}^i_n (x)\hat{T}^{m-2-i}_n(y)\hat{S}_n(x,y)+ \sum_{ i=0}^{m-3}\sum_{j=0}^{m-3-i} x^i y^j \hat{T}^i_n(x) \hat{T}^j_n(y)\hat{S}_n(x,1) \hat{S}_n(1,y)\right)\nonumber
\label{gen_rmm}
\eea
\end{widetext}
where  $T_{0}(x)=1-\hat{S}_{0}(x,y) =p$.
Notice that the Eqs.\ $(\ref{gen_rmm})$ reduce to the Eqs.\ $(\ref{gen_m})$ for a deterministic hyperbolic manifold formed by $m'$ polygons for  $q_{m^{\prime}}=\delta_{m,m^{\prime}}$.

\begin{figure}
\begin{center}
\includegraphics[width=\columnwidth]{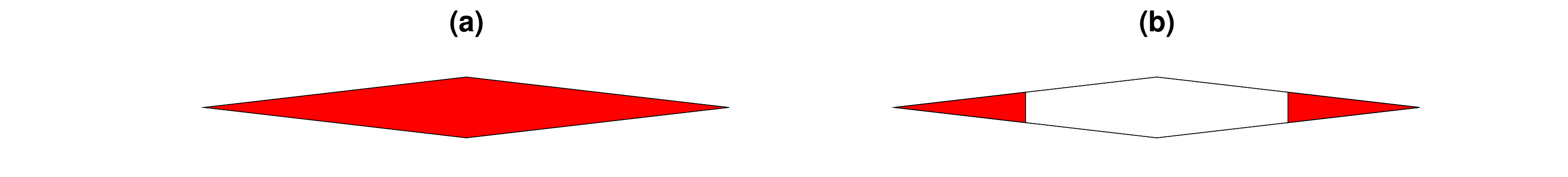}
\caption{Diagramatic representation of generating functions $\hat{T}_n(x)$ (a) and $\hat{S}_n(x,y)$ (b). Filled areas indicate clusters that either connect ($\hat{T}_n(x)$) or do not connect ($\hat{S}_n(x,y)$) the endnodes.}
\label{fig:diagrams1}
\end{center}
\end{figure}

\begin{figure}
\begin{center}
\includegraphics[width=\columnwidth]{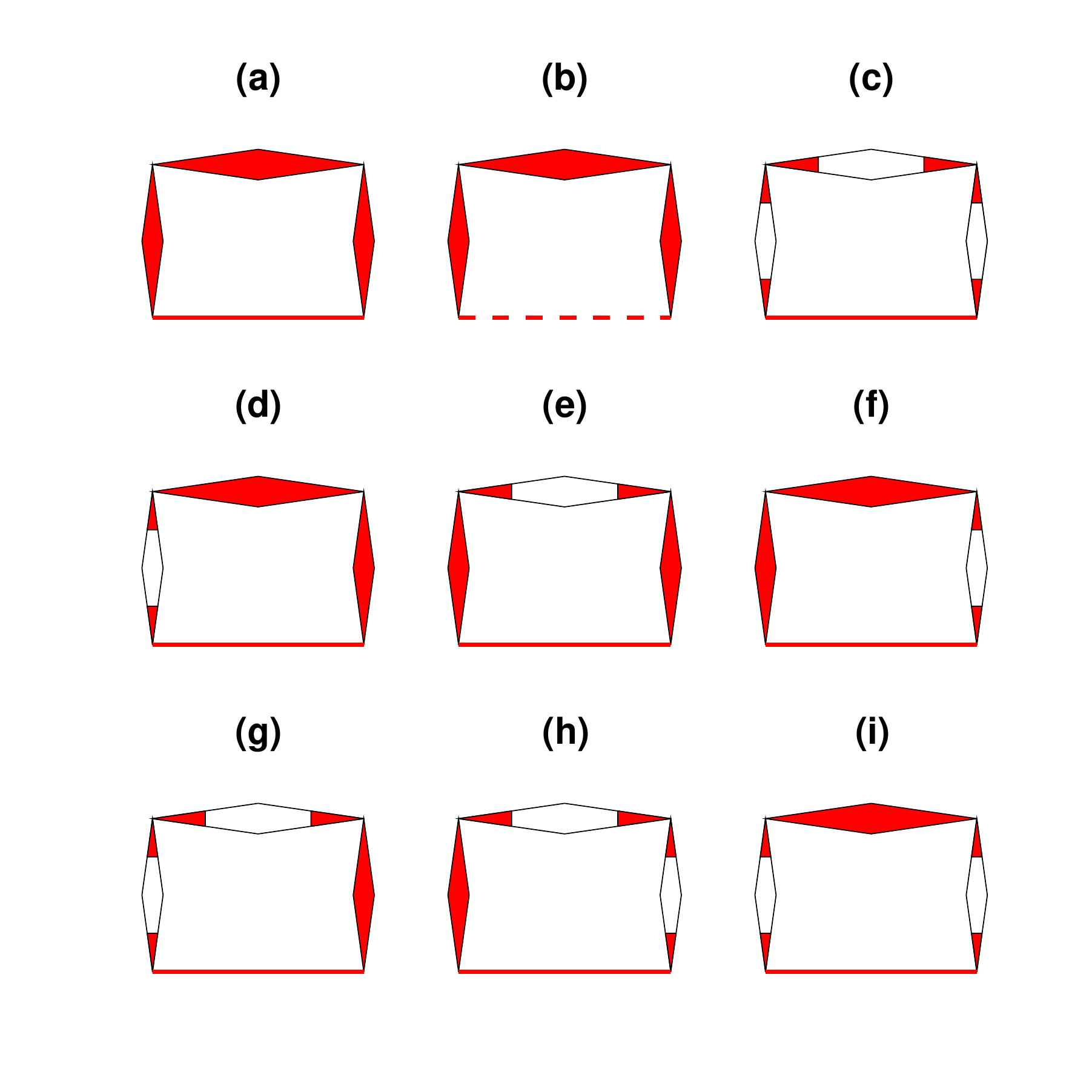}
\caption{The diagrams contributing to $\hat{T}_{n+1}(x)$ for the deterministic manifold with $m=4$ are shown. The contributions from the configurations (a), (b) and (d)-(h) are: $px^2\hat{T}_n^3(x)$ (a), $(1-p) x^2\hat{T}_n^3(x)$ (b),  $p x^2\hat{T}_n^2(x)\hat{S}_n(x,x)$ (d,e,f), $p x\hat{T}_n(x)\hat{S}_n^2(x,1)$ (g,h). The comprehensive contribution of configurations (c) and (i) is $\hat{S}_n^2(x,1)$.}
\label{fig:diagrams2}
\end{center}
\end{figure}
\begin{figure}
\begin{center}
\includegraphics[width=\columnwidth]{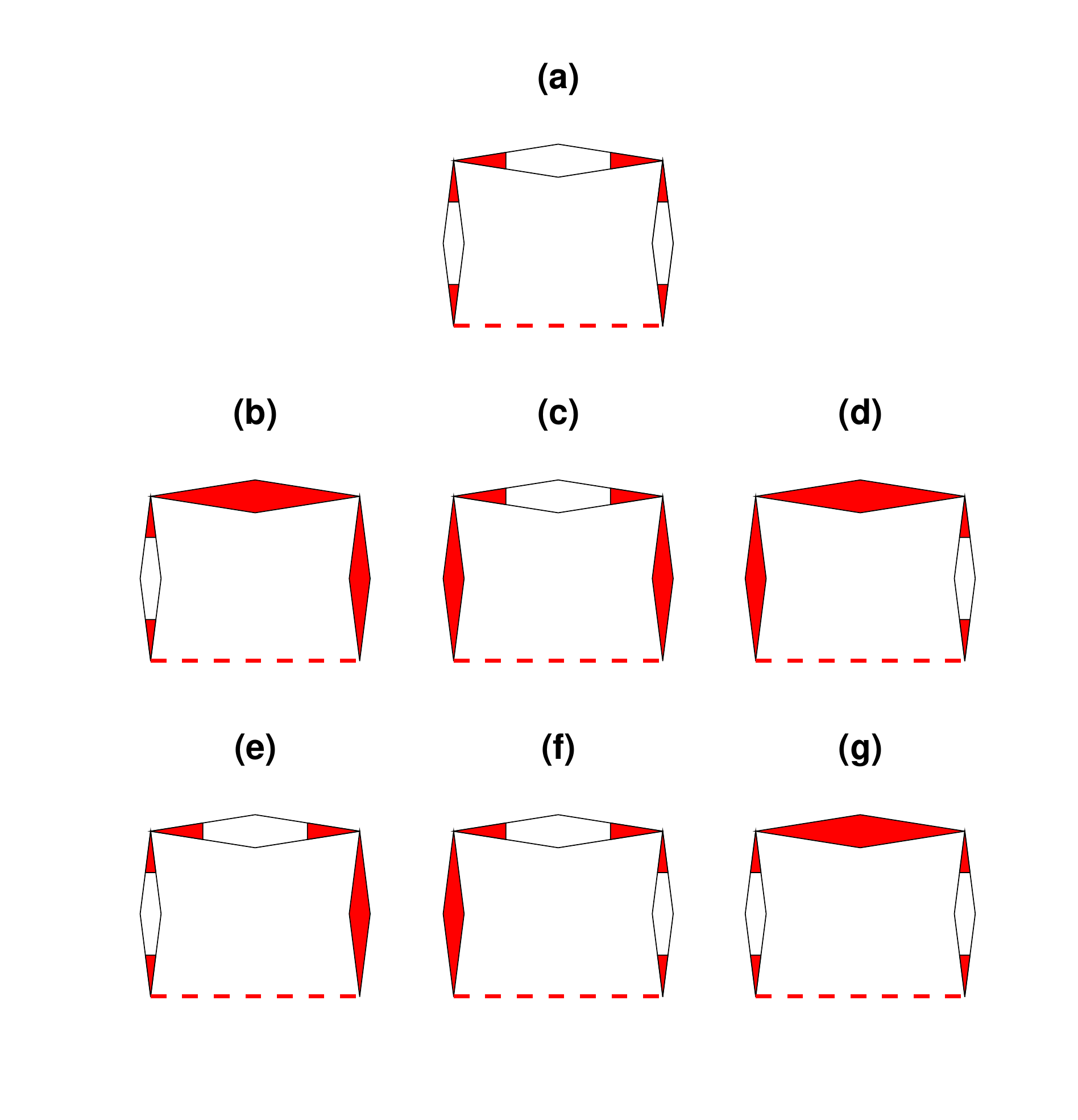}
\caption{The diagrams contributing to $\hat{S}_{n+1}(x)$ for the deterministic manifold with $m=4$ are shown. The comprehensive contribution from the configurations (a) and (g) is $(1-p)\hat{S}_n(x,1)\hat{S}_n(y,1)$. The contribution form the other configurations are:
  $(1-p)y^2\hat{T}_n^2(y)\hat{S}_n(x,y)$ (b), $(1-p) xy\hat{T}_n(x)\hat{T}_n(y)\hat{S}_n(x,y)$ (c),  $(1-p)x^2\hat{T}_n^2(x)\hat{S}_n(x,y)$ (d), $(1-p) y\hat{T}_n(y)\hat{S}_n(x,1)\hat{S}_n(y,1)$ (e),$(1-p) x\hat{T}_n(x)\hat{S}_n(x,1)\hat{S}_n(y,1)$ (f). }
\label{fig:diagrams3}
\end{center}
\end{figure}
\section{Fractal exponent}
\subsection{General derivation}

In this section we investigate how the expected size $R_n$ of the giant component connected to the two initial nodes grows with the number of generations $n$ and from this expression we will derive the fractal exponent $\psi$.
The expected size of the giant component $R_n$ can be derived from the generating function $\hat{T}_n(x)$ by differentiation, i.\,e.,
\bea
R_n=\left.\frac{d\hat{T}_n(x)}{dx}\right|_{x=1}.
\eea
To this end we rewrite Eqs.\ $(\ref{gen_rmm})$ in terms of the vector  
\bea
\mathbf V_n(x)&=& \left(V_n^1(x),V_n^2(x),V_n^3(x)\right)^\top\nonumber \\ &=&\left(\hat{T}_n(x),\Sigma_n(x),S_n(x)\right)^\top,\eea where $\Sigma_n(x)=\hat{S}_n(x,x)$ and $S_n(x)=\hat{S}(1,x)$, obtaining a recursive equation of the type
\bea\label{eq:poly:V}
\mathbf V_{n+1}(x) =  \mathbf F_n(\mathbf V_{n}(x), x ).
\eea
This system of equations can be differentiated obtaining
\bea
\frac{d{\bf V}_{n+1}(x)}{dx}=\sum_{s=1}^3\frac{\partial{\bf F}_n}{\partial{V}_n^{s}(x)}\frac{d{ V}_n^{s}(x)}{dx}+\frac{\partial{\bf F}_n}{\partial x},
\label{Vprime}
\eea
with initial condition ${\bf V}^{\prime}_0=(0,0,0)$ (the initial nodes are not counted).
Since the non-homogeneous term ${\partial{\bf F}_n}/{\partial x}$  is subleading with respect to the homogeneous one,  for $n\gg 1$ and $T<1$ we 
have:{ 
\bea
 \dot{ \mathbf V}_{n+1} \simeq {\mathcal D}_n \prod_{n'=1}^n\lambda_{n'}{\bf u}_n,
\eea
where $\lambda_n$ and ${\bf u}_n$ are the largest eigenvalue and  the the corresponding eigenvector of the Jacobian matrix  ${\mathbf J}_n$ defined as
\bea
\left[J_n\right]_{ij}=\left.\frac{\partial{F}^{i}(x)}{\partial{V}^{j}(x)}\right|_{{\bf V}(x)={\bf V}_{n}(1); x=1},
\eea
and ${\mathcal D}_n$ is given by 
\bea
{\mathcal D}_n= \left(\prod_{n'=2}^n \braket{{\bf u}_{n'}|{\bf u}_{n'-1}} \right)\braket{{\bf u}_1|\dot{\bf V}_0},
\eea
with  $\dot{\bf V}_0={\partial{\bf F}_0}/{\partial x}$.
Assuming that for $p\simeq p_c$, ${\mathcal D}_n$ is in first approximation independent of $n$, it}  follows that $R_n=\dot{V}_n^1$ scales like
\bea
R_n\sim \prod_{n'=1}^n\lambda_{n'}=\exp\left[{\sum_{n'=0}^n \ln \lambda_n}\right].
\label{Rna}
\eea

By defining  $\psi_n$ as 

\bea
\psi_n=\frac{\ln \lambda_n}{\ln ( {\avg{m}-1)}}
\label{psi_n}
\eea

\noindent and using the expression for $\bar{N}_n$ given by Eq.\ $(\ref{barN})$ we get for $n\gg 1$
\bea
R_n\sim \bar{N}_n^{\psi}
\eea
where the fractal exponent $\psi$ is given by 
\bea
\psi=\lim_{n\to \infty} \psi_n.
\label{lim_psi}
\eea

\subsection{Deterministic and random hyperbolic manifolds}

Here we will perform the explicit calculation outlined in the previous subsection by treating explicitly the random hyperbolic manifolds. In fact  the results for the deterministic hyperbolic manifolds can be deduced from this calculations by considering a distribution $q_m$ equal to a Kronecker delta.
The recursive equations for $\hat{T}_n(x),\Sigma_n(x)$ and $S_n(x)$ can be directly deduced from Eqs.\ (\ref{gen_rmm}) {by setting $y=x$ and $y=1$ and using $\hat{T}_n(1)=1-\Sigma_n(1,1)=1-S_n(1)=T_n$ and read 
\begin{widetext}
\bea
\hat{T}_{n+1}(x) &=&\sum\limits_{m=3}^\infty q_m \left[x^{m-2}\hat{T}_n^{m-1}(x)+ p(m-1)  x^{m-2}   \hat{T}^{m-2}_n(x)\Sigma_n(x)+ p\left(\sum_{i=0}^{m-3}(i+1) x^i \hat{T}_n^i(x) \right)S_n^2(x)\right],\nonumber \\
\Sigma_{n+1}(x) &=&(1-p)\sum\limits_{m=3}^\infty q_m\left[(m-1)x^{m-2} \hat{T}_n^{m-2}(x) \Sigma_n(x) +\left(\sum_{i=0}^{m-3}(i+1)x^i \hat{T}_n^i(x)\right) S^2_n(x) \right],\nonumber \\
S_{n+1}(x) &=& (1-p)\sum\limits_{m=3}^\infty q_m\left(\sum_{i=0}^{m-2}x^i\hat{T}_n^i(x)\right)S_n(x).
\label{eq:poly:det}
\eea
\end{widetext}
By differentiating Eqs.\ \eqref{eq:poly:det} with respect to $\mathbf{V}_n$ and putting $x=1$, and using $\hat{T}_n(1)=1-\Sigma_n(1,1)=1-S_n(1)=T_n$, we get the Jacobian $\mathbf{J}_n$.
In particular by using the  mathematical relation
\bea
(1-T_n)\sum_{i=0}^{m-3} i(i+1)T_n^{i-1}&=&2\sum_{i=0}^{m-3}(i+1)T_n^i\nonumber \\
&&-(m-1)(m-2)T_n^{m-3}\nonumber \\
\eea
and
\bea
(1-T_n)\sum_{i=0}^{m-3}(i+1)T_n^i&=&\sum_{i=0}^{m-2}T_n^i-(m-1)T_n^{m-2}\nonumber
\eea
one can show that the Jacobian ${\mathbf J}_n$ can be expressed as
\begin{widetext}
\bea
{\mathbf J}_n=\left(
\begin{array}{ccc}
 Q'(T_n)+2p \left[\text{H}(T_n)-Q'(T_n)\right] & p Q'(T_n) & 2 p \left[\text{H}(T_n)-Q'(T_n)\right] \\
 2 (1-p) \left[\text{H}(T_n)-Q'(T_n)\right] & (1-p) Q'(T_n) & 2 (1-p) \left[\text{H}(T_n)-Q'(T_n)\right] \\
 (1-p) \left[\text{H}(T_n)-Q'(T_n)\right] & 0 &(1-p)\text{H}(T_n) \\
\end{array}
\right),
\eea
\end{widetext}
where  $Q(T)$  is defined in Eq.\ (\ref{QT}) and $H(T)$ is defined as
\bea
H(T)=\sum_{m=3}^{\infty}q_m \sum_{i=0}^{m-2} T^i,
\eea
which admits for  $T<1$ the expression
\bea
H(T)=\frac{1-Q(T)}{1-T}.
\eea
Similarly  it can be shown that ${\partial{\bf F}_n}/{\partial x}$ is given by 
\bea
\frac{\partial{\bf F}_n}{\partial x}=\left(\begin{array}{c} pT_n2[H(T_n)-Q^{\prime}(T_n)]+[T_nQ^{\prime}(T_n)-Q(T_n)]\\2(1-p)T_n[H(T_n)-Q^{\prime}(T_n)]\\(1-p)T_n[H(T_n)-Q^{\prime}(T_n)]\end{array}\right).\nonumber
\eea
For  $T_n<1$  the Jacobian ${\mathbf{J}}_n$ has the largest eigenvalue $\lambda_n$ given by
\bea
\lambda_n&=&\frac{1}{2} \left[\sqrt{\hat{\Delta}(T_n)}+ \hat{K}(T_n) \right],
\label{lambda_n}
\eea
where $\hat{\Delta}(T_n)$ and $\hat{K}(T_n)$ are  given by 
\bea
\hat{\Delta}(T_n)&=&\left[ \hat{K}(T_n)\right]^2-4  (1-{p})H(T_n) Q'(T_n),\nonumber \\
\hat{K}(T_n)&=&({p}+1)H(T_n)+ (1-2 {p})Q'(T_n).
\eea
For  $T_n=1$, instead, the largest eigenvalue is given by
{$$\lambda_n=\Avg{m}-1.$$
{ 
The eigenvector ${\bf u}_n$ corresponding to the largest eigenvalue is 
\bea
{\bf u}_n={\mathcal C}\left(\begin{array}{c}\hat{K}(T_n)-2(1-p)H(T_n)+\sqrt{\hat{\Delta}(T_n)}\\
4(1-p)[H(T_n)-Q^{\prime}(T_n)] \\ 
2(1-p)[H(T_n)-Q^{\prime}(T_n)]
\end{array}\right),
\eea
where ${\mathcal C}$ is the normalization constant.
For  $T_n=1$ and $p=p_c$,  the  eigenvector ${\bf u}_n$ is given by
$${\bf u}_n=\left(1,0,0\right)^{\top}.$$}
In Figure $\ref{fig:psifig}$ we show the fractal exponent $\psi_n$ as a function of $p$ for deterministic hyperbolic manifolds with $m=3,4,5$ and for $n=1000$.
\begin{figure}
\includegraphics[width=1\columnwidth]{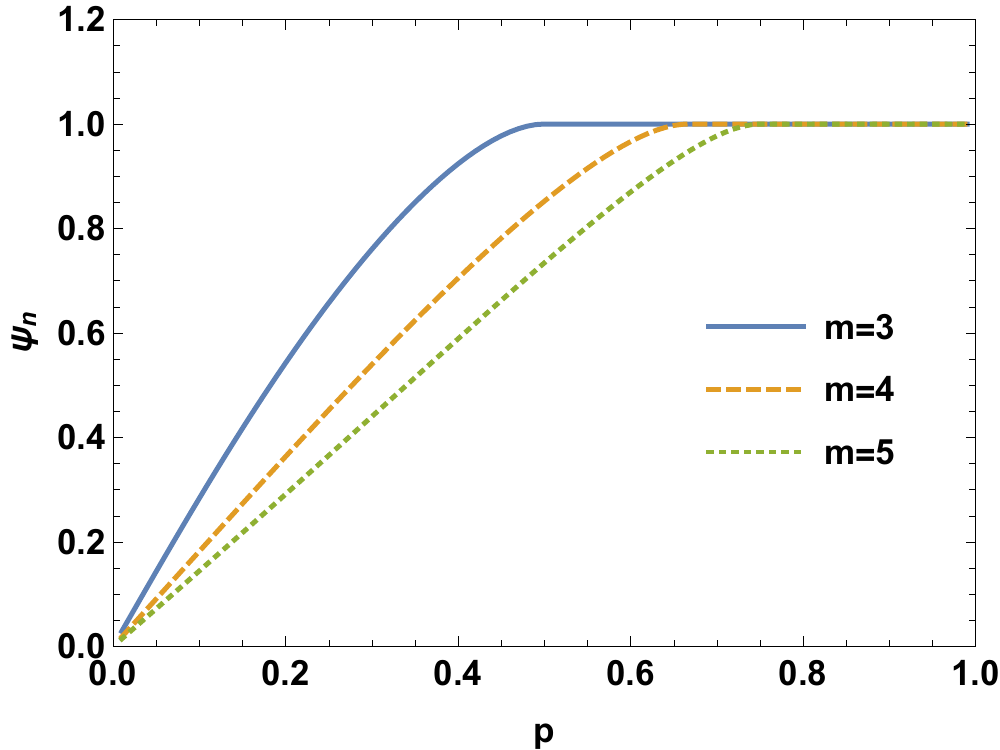}
\caption{The  exponent $\psi_n$ is shown as a function of $p$ for the deterministic hyperbolic manifold with $m=3,4,5$. All curves are obtained for  $n=1000$.}
\label{fig:psifig}
\end{figure}

\begin{table}[htb!]
\bgroup
\def\arraystretch{1.5}
\begin{center}
\begin{tabular}{ccc}
$\gamma$\hspace{0.5cm} 		&   Expansion of $1-\psi_n$& Expansion of $1-\psi $\hspace{0.5cm}		 \\
\hline
$\gamma>4$	&   	$\hat{A}_{\gamma}\left(\Delta T_n\right)^2$						&$\tilde{A}_{\gamma}\left(\Delta p\right)^2$	\\
$\gamma=4$	&   	$\hat{A}_{\gamma}\left(\Delta T_n\right)^2\ln |\Delta T_n|$						&$\tilde{A}_{\gamma}\left(\Delta p\right)^2\ln |\Delta p|$	\\
$\gamma\in (3,4)$	&   	$\hat{A}_{\gamma}\left|\Delta T_n\right|^{\gamma-2}$						&$\tilde{A}_{\gamma}\left|\Delta p\right|^{\gamma-2}$	\\
$\gamma=3$		&   	$\hat{A}_{\gamma}\Delta T_n\ln|\Delta T_n|$		& $\tilde{A}_{\gamma}\Delta p[\ln|\Delta p|]^2$	\\
$\gamma\in (2,3)$	&  	$\hat{A}_{\gamma} |\Delta T_n|^{{\gamma-2}}$				&$\tilde{A}_{\gamma} |\Delta p|$		
\end{tabular}\\
\end{center}
\caption{
 Asymptotic expansion of $1-\psi_n$  and $1-\psi $ in $\Delta T_n=T_n-1$ and in $\Delta p=p-p_c$ valid for $\Delta p\ll 1$ and $|\Delta T_n|\ll 1$. Here, we consider random hyperbolic manifolds with  asymptotic power-law scaling of the $q_m$ distribution, i.\,e.,  $q_m\simeq C m^{-\gamma}$ for $m\gg 1$.}
\label{table:expansion_psi}
\egroup
\end{table}%
\subsection{Scaling of the fractal exponent}
\label{sec_psi}
The fractal exponent $\psi$ is defined as the limit for $n\to \infty$ of $\psi_n$ related to the maximum eigenvalue $\lambda_n$ of the Jacobian matrix ${\mathbf J}_n$ by  Eq.\ (\ref{psi_n}).
Since $\lambda_n$ is expressed in Eq.\ (\ref{lambda_n}) in terms of $Q'(T_n)$ and $H(T_n)$, by  expanding these latter quantities  for $0<T_c-T_n\ll 1$ and $0<p_c-p\ll 1$ we can derive the critical behavior of $\psi_n$
that can be expressed as  a function of $\Delta T_n=T_n-1$, i.\,e.,
\bea
\psi_n=f(\Delta T_n).
\label{psi_n_scaling}
\eea
From this scaling, by performing the limit $n\to \infty$, we can derive the critical scaling of the fractal exponent.
In the  following we will analyse the  critical behaviour of $\psi$ for distributions $q_m$ with convergent first, second and third moments $\avg{m},\avg{m^2}$ and $\avg{m^3}$ and with asymptotic power-law scaling $q_{m}\simeq C m^{-\gamma}$ with $\gamma>2$.  In particular we will show that for hyperbolic manifolds with scale-free distribution $q_m$ the critical behavior becomes dependent on the exponent $\gamma$ as summarized in Table $\ref{table:expansion_psi}$.

\subsubsection{Case with converging $\avg{m},\avg{m^2}$ and $\avg{m^3}$}
The first case we consider is the case of a generic distribution $q_m$ having convergent first, second and third moments $\avg{m},\avg{m^2}$ and $\avg{m^3}$.
By expanding $Q'(T_n)$ and $H(T_n)$ close to the critical point, i.\,e., for  $p=p_c+\Delta p$,  and  $T_n=T_c+\Delta T_n$  for $\Delta p<0$ and $\Delta T_n<0$ but small in absolute values, i.\,e., $|\Delta T_n|\ll1$ and $|\Delta p|\ll1$, we obtain
\bea
Q'(T_n)&=&(\avg{m}-1)+\Avg{(m-1)(m-2)}\Delta T_n+\nonumber \\&&\hspace{-7mm}\frac{1}{2}\Avg{(m-1)(m-2)(m-3)}(\Delta T_n)^2+o((\Delta T_n)^2)\nonumber \\
H(T_n)&=&(\avg{m}-1)+\frac{1}{2}\Avg{(m-1)(m-2)}\Delta T_n\nonumber \\&&\hspace{-7mm}+\frac{1}{6}\Avg{(m-1)(m-2)(m-3)}(\Delta T_n)^2+o((\Delta T_n)^2)\nonumber
\eea
Therefore by using the definition of $\psi_n$ (Eq.\ (\ref{psi_n})) and the explicit expression of $\lambda_n$ (Eq.(\ref{lambda_n})) we can derive the scaling of $\psi_n$ as a function of $\Delta T_n$ 
\bea
\psi_n= 1-\hat{A} (\Delta T_n)^2+o((\Delta T_n)^2)\nonumber
\label{qpsin}
\eea
where the constant $\hat{A}$ is given by 
\begin{widetext}
\bea
\hat{A}=\frac{3\Avg{(m-1)(m-2)}+(\avg{m}-1)(\Avg{m}-2)(\Avg{(m-1)(m-2)(m-3)})}{6(\avg{m}-1)^2(\avg{m}-2)\ln (\Avg{m}-1)}.
\label{hA}
\eea
\end{widetext}
Specifically in the case in of a deterministic hyperbolic manifold, when $q_{m'}=\delta_{m',m}$, we have $\hat{A}=\hat{A}_m$ given by 
\bea
\hat{A}_m=\frac{1}{6}\frac{m(m-2)}{\ln(m-1)}.
\label{hAm}
\eea
In order to derive the scaling of the fractal exponent $\psi$ with $\Delta p$  we use the fact that for  $p_c-p\ll 1$  and $n\to \infty$ we have that $\Delta T=T-T_c$ is proportional to $\Delta p$ with proportionality constant  $A$ (see Eq.\ (\ref{eq:polyQ:expansion})). Therefore we predict that close to the critical point, for $0<\Delta p\ll1$, $\psi$ is given by
  \bea
  \psi = 1- \tilde{A} ( p_c - p)^2+o((p_c-p)^2)
  \eea where
  \begin{widetext}
 \bea
\tilde{A}=\frac{2(\avg{m}-1)^2}{3(\Avg{(m-1)(m-2)})^2}\frac{3\Avg{(m-1)(m-2)}+(\avg{m}-1)(\Avg{m}-2)(\Avg{(m-1)(m-2)(m-3)})}{(\avg{m}-2)\ln (\Avg{m}-1)}
\label{tA}
\eea
\end{widetext}
For a deterministic hyperbolic manifold with $q_{m'}=\delta_{m',m}$ with $m\geq 3$
the constant  $\tilde{A}$ equals $\tilde{A}_m$,  where 
\bea
  \tilde{A}_m=\frac{2m (m-1)^2 }{3 (m-2) \ln (m-1)}.
\eea
Interestingly, $\tilde{A}_m$ gains its minimum value at $m=4$,  which indicates that among this class of hyperbolic manifolds, the square hyperbolic manifold features the slowest conversion  convergence of $\psi$ to 1.
\subsubsection{Case with power-law distribution $q_m$ with power-law exponent $\gamma$}
Here we consider the scaling of $\psi_n$ as a function of $\Delta T_n$ and the scaling of the fractal exponent $\psi$ as a function of $\Delta p$ for random hyperbolic manifolds with asymptotic power-law distribution $q_m\simeq Cm^{-\gamma}$ as a function of the  the value of the power-law exponent $\gamma$.
\begin{itemize}
\item[(i)]{\em Case $\gamma>4$}\\
This case reduces to the case of the generic distribution $q_m$ having finite first, second, and third moments $\avg{m},\avg{m^2}$ and $\avg{m^3}$ convergent, studied in the precedent subsection.
In fact in this case we can expand the functions $Q'(T)$ and $H(T)$ for $\Delta p<0$ and $\Delta T_n<0$, with $|\Delta T_n|\ll1$ and $|\Delta p|\ll1$,  getting 
\bea
Q'(T_n)&=&(\avg{m}-1)+\Avg{(m-1)(m-2)}\Delta T_n+\nonumber \\&&\hspace{-7mm}c_{\gamma}(\Delta T_n)^2+o((\Delta T_n)^2)\nonumber \\
H(T_n)&=&(\avg{m}-1)+\frac{1}{2}\Avg{(m-1)(m-2)}\Delta T_n\nonumber \\&&\hspace{-7mm}+d_{\gamma}(\Delta T_n)^2+o((\Delta T_n)^2),\nonumber
\eea
where $c_{\gamma}=\Avg{(m-1)(m-2)(m-3)}/2$ and $d_{\gamma}=\Avg{(m-1)(m-2)(m-3)}/6$  (see Table $\ref{table:acd}$ for their expression in the case of a pure power-law distribution $q_m$).
Using these expressions in Eq.(\ref{lambda_n}) for $\lambda_n$ we derive  the critical scaling of  $\psi_n$ given by 
\bea
\psi_n=1-\hat{A}_{\gamma}(\Delta T_n)^2+o((\Delta T_n)^2),
\eea
with $\hat{A}_{\gamma}=\hat{A}$ given by Eq.\ (\ref{hA}) .
Moreover in the limit $n\to \infty$ we predict that the fractal exponent $\psi$ has critical behavior
\bea
\psi=1-\tilde{A}_{\gamma}(\Delta p)^2+o((\Delta p)^2),
\eea
with $\tilde{A}_{\gamma}=\tilde{A}$ given by Eq.\ (\ref{tA}).
\item[(ii)]{\em Case $\gamma=4$}\\
In this case we can perform the asymptotic expansion of  $Q'(T_n)$ and $H(T_n)$ for $|\Delta T_n|\ll 1$ for $\Delta p<0$ and $\Delta T_n<0$ with $|\Delta T_n|\ll1$ and $|\Delta p|\ll1$ obtaining 
\bea
Q'(T_n)&=&(\avg{m}-1)+\Avg{(m-1)(m-2)}\Delta T_n\nonumber \\
&&+c_{\gamma} (\Delta T_n)^2 \ln |\Delta T_n|+{\mathcal O}((\Delta T_n)^2),\nonumber \\
H(T_n)&=&(\avg{m}-1)+\frac{1}{2}\Avg{(m-1)(m-2)}\Delta T_n\nonumber \\
&&+d_{\gamma}(\Delta T_n)^2 \ln |\Delta T_n|+{\mathcal O}((\Delta T_n)^2),\eea
where $c_{\gamma}$ and $d_{\gamma}$ are constants (see Table $\ref{table:acd}$ for their expression in the case of a pure power-law distribution $q_m$).
By inserting these asymptotic expansions in  $\lambda_n$ (given by Eq.(\ref{lambda_n})) we can derive the asymptotic expansion of $\psi_n$
\bea
\psi_n= 1-\hat{A}_{\gamma}(\Delta T_n)^2 \ln |\Delta T_n|+{\mathcal O}(|\Delta T_n|^2)
\eea
where
\bea
\hat{A}_{\gamma}=\frac{c_{\gamma}-2d_{\gamma}}{(\avg{m}-1)\ln (\avg{m}-1)}.
\label{Ag}
\eea
By performing the limit  $n\to \infty$ and using the fact that close to the upper percolation threshold,  Eq.\ $(\ref{Tmf})$ holds we can easily show that the fractal exponent $\psi$ scales like
\bea
\psi= 1-\tilde{A}_{\gamma}(\Delta p)^2 \ln |\Delta p|+{\mathcal O}(|\Delta p|^2),
\eea
where $\tilde{A}_{\gamma}=\hat{A}_{\gamma} A_{\gamma}^{2}$.
\item[(iii)]{\em Case  $\gamma\in (3,4)$}\\
In this case we consider the asymptotic expansion of $Q'(T_n)$ and $H(T_n)$ for $\Delta p<0$ and $\Delta T_n<0$ with $|\Delta T_n|\ll1$ and $|\Delta p|\ll1$  given by
\bea
Q'(T_n)&= &(\avg{m}-1)+\Avg{(m-1)(m-2)}\Delta T_n\nonumber \\
&&+c_{\gamma}|\Delta T_n|^{\gamma-2}+o(|\Delta T_n|^{\gamma-2},\nonumber \\
H(T_n)&=&(\avg{m}-1)+\frac{1}{2}\Avg{(m-1)(m-2)}\Delta T_n\nonumber \\
&&+d_{\gamma}|\Delta T_n|^{\gamma-2}+o(|\Delta T_n|^{\gamma-2}),\nonumber 
\eea
where $c_{\gamma}$ and $d_{\gamma}$ are constants (see Table $\ref{table:acd}$ for their expression in the case of a pure power-law distribution $q_m$).
By following the same procedure applied to previous case we obtain the asymptotic scaling for $\psi_n$ given by
\bea
\psi_n=1-\hat{A}_{\gamma}|\Delta T_n|^{\gamma-2}+o(|\Delta T_n|^{\gamma-2}),
\eea
where $A_{\gamma} $ is given by Eq.(\ref{Ag}).

Finally performing the limit $n\to \infty$ and using Eq.\ (\ref{Tmf}) we derive the following asymptotic expansion for the fractal exponent $\psi$,
\bea
\psi=1-\tilde{A}_{\gamma}|\Delta p|^{\gamma-2}+o(|\Delta p|^{\gamma-2})
\eea
where $\tilde{A}_{\gamma}=\hat{A}_{\gamma} A_{\gamma}^{(\gamma-2)}$.

\item[(iv)]{\em Case  $\gamma=3$}\\
For $\gamma=3$ we consider the asymptotic expansion of  $Q'(T)$ and $H(T)$ for $\Delta p<0$ and $\Delta T_n<0$ with  $|\Delta T_n|\ll1$ and $|\Delta p|\ll1$,  which is given by 
\bea
Q'(T_n)&=&(\avg{m}-1)+c_{\gamma} \Delta T_n \ln |\Delta T_n|+{\mathcal O}(\Delta T_n),\nonumber \\
H(T_n)&=&(\avg{m}-1)+d_{\gamma} \Delta T_n \ln |\Delta T_n|+{\mathcal O}(\Delta T_n),\nonumber \eea
where $c_{\gamma}$ and $d_{\gamma}$ are constants (see Table $\ref{table:acd}$ for their expression in the case of a pure power-law distribution $q_m$). Using this expressions in the definition of  $\lambda_n$ given by Eq.(\ref{lambda_n}) we derive the critical scaling of $\psi_n$ given by 
\bea
\psi_n=1-\hat{A}_{\gamma}\Delta T_n \ln |\Delta T_n|+{\mathcal O}(\Delta T_n)
\eea
where $A_{\gamma} $ is given by Eq.(\ref{Ag}).

By performing the limit of $\psi_n$ for $n\to\infty$ and using the scaling relation determined in Eq.\ (\ref{T3}) we obtain
\bea
\psi=1-\tilde{A}_{\gamma}\Delta p \left[\ln |\Delta p|\right]^2+o((\Delta p)\left[\ln |\Delta p|\right]^2)
\eea
\item[(v)]{\em Case  $2<\gamma<3$}\\
Finally in this case the asymptotic expansion of $Q'(T_n)$ and $H(T_n)$ for $\Delta p<0$ and $\Delta T_n<0$, $|\Delta T_n|\ll1$ and $|\Delta p|\ll1$ is given by 
\bea
Q'(T_n)&=&(\avg{m}-1)+c_{\gamma}|\Delta T_n|^{\gamma-2}+o(|\Delta T_n|^{\gamma-2}),\nonumber \\
H(T_n)&=&(\avg{m}-1)+d_{\gamma}|\Delta T_n|^{\gamma-2}+o(|\Delta T_n|^{\gamma-2}),\nonumber 
\eea
where $c_{\gamma}$ and $d_{\gamma}$ are constants (see Table $\ref{table:acd}$ for their expression in the case of a pure power-law distribution $q_m$).
These expression  leads to the critical scaling of $\psi_n$ given by  
\bea
\psi_n=1-\hat{A}_{\gamma}|\Delta T_n|^{\gamma-2}+o(|\Delta T_n|^{\gamma-2})
\eea
 where $A_{\gamma} $ is given by Eq.(\ref{Ag}).
By performing the limit of $\psi_n$ for $n\to\infty$ and using the scaling relation determined in Eq.\ (\ref{T23}) with $\beta=1/(\gamma-2)$, we obtain the critical behavior of the fractal exponent $\psi$, i.\,e., 
\bea
\psi=1-\tilde{A}_{\gamma}|\Delta p|+o(|\Delta p|)
\eea
where $\tilde{A}_{\gamma}=\hat{A}_{\gamma}A_{\gamma}^{\gamma-2}$
\end{itemize}
{Thus, for a distribution $q_m$ that scales asymptotically as a power-law, we find an anomalous exponent $\psi$ that depends upon the scaling exponent $\gamma$ as described in Table $\ref{table:expansion_psi}$.}

\begin{table*}
\bgroup
\def\arraystretch{1.5}
\begin{center}
\begin{tabular}{cccc}
$\gamma$		&   $a_{\gamma}$ &$c_{\gamma}$ & $d_{\gamma}$\hspace{0.5cm}		 \\
\hline
$\gamma\geq 4$	&  $\frac{\zeta(\gamma-2)-3\zeta(\gamma-1)+2\zeta(\gamma)}{2(-1-2^{-\gamma}+\zeta(\gamma))}$ &	$\frac{360}{765-8\pi^4}$			& $\frac{120}{765-8\pi^4}$	\\
$\gamma\in(3,4)$		&  $\frac{\zeta(\gamma-2)-3\zeta(\gamma-1)+2\zeta(\gamma)}{2(-1-2^{-\gamma}+\zeta(\gamma))}$ &	$\frac{2^{\gamma}(\gamma-1)\Gamma[1-\gamma]}{-1-2^{\gamma}+2^{\gamma}\zeta(\gamma)}$		& $\frac{2^{\gamma}\Gamma[1-\gamma]}{1+2^{\gamma}-2^{\gamma}\zeta(\gamma)}$	\\
$\gamma=3$	&  $\frac{4}{9-8\zeta(3)}$ &	$\frac{4}{9-8\zeta(3)}$				&$\frac{4}{9-8\zeta(3)}$		\\
	$\gamma\in(2,3)$		& $\frac{2^{\gamma}\Gamma[1-\gamma]}{-1-2^{\gamma}+2^{\gamma}\zeta(\gamma)}$  &	$\frac{2^{\gamma}(\gamma-1)\Gamma[1-\gamma]}{-1-2^{\gamma}+2^{\gamma}\zeta(\gamma)}$		& $\frac{2^{\gamma}\Gamma[1-\gamma]}{1+2^{\gamma}-2^{\gamma}\zeta(\gamma)}$	
\end{tabular}\\
\end{center}
\caption{
Values of $a_{\gamma}$, $c_{\gamma}$ and $d_{\gamma}$ determining respectively the asymptotic expansion of $Q(T), Q^{\prime}(T)$ and $H(T)$ for a pure power-law $q_m$ distribution given by  $q_m= C m^{-\gamma}$, for $m\geq 3$.}
\label{table:acd}
\egroup
\end{table*}%
\section{Fraction of nodes in the giant component $P_{\infty}$}
\subsection{General framework}
In this section our goal is to investigate the nature of the percolation phase transition at the upper critical threshold $p_c$ using RG arguments already introduced in Ref.\ \cite{RG}. For this transition the order parameter is given by the fraction $P_{\infty}$ of nodes in the giant component in an infinite network, i.\,e.,
\bea
P_{\infty}&=&\lim_{n\to \infty}\frac{R_n}{\bar{N}_n},
\eea
By using Eq.\ (\ref{Rna}) for approximating $R_n$ when $n\gg1$ we obtain
\bea
P_{\infty}&\simeq&\lim_{n\to \infty}\frac{1}{\bar{N}^{(0)}_n}\prod_{n'=1}^{n}\lambda_{n'}\nonumber \\
&\simeq &\exp\left[-\ln[\avg{m-1}]\int_0^{\infty}dn(1-\psi_n)\right],
\label{pinft}
\eea
where the last expression is derived by using a continuous approximation for $n$.
Therefore  in order to evaluate the critical behavior $P_{\infty}$ we need to know the dependence of $\psi_n$ on $n$. 
Close to the percolation threshold, for $0<p_c-p \ll 1$ we can use Eq.
$(\ref{psi_n_scaling})$ and its explicit expression derived in Sec. $\ref{sec_psi}$ in order to express $\psi_n$ as a function of 
$\Delta T_n$.
Finally the dependence of $\Delta T_n$ on $n$ can be derived by expanding the RG Eq.\ $(\ref{RG})$ close to the upper percolation threshold $p_c$.
In the following we will  derive the critical behavior of the order parameter $P_{\infty}$ by following the inverse order. Firstly we will derive the functional behavior of $\Delta T_n$ on $n$, then we will use the scaling of $\psi_n$ as a function of $\Delta T_n$ to predict the nature of the percolation transition at $p=p_c$ in the continuous approximation.
We will consider first the cases in which the second moment  $\avg{m^2}$ is convergent, finding that the percolation transition is always discontinuous. Subsequently we will investigate the cases for which $q_m$ has an asymptotic power-law decay $q_{m}\simeq Cm^{-\gamma}$ for $m\gg 1$.  In this case  we observe as a function of $\gamma$ different universality classes  summarized in Table $\ref{table:pinf}$. As long as both $\avg{m}$ and $\avg{m^2}$ are convergent we find always discontinuous  transitions although the universality classes as a function of $\gamma$ varies determining different critical scalings. When $\avg{m^2}$  is divergent we find instead  that  the percolation transition becomes continuous.
Our approach  uses the continuous approximation, and to validate this approach  our analytical results are compared with exact numerical results performed for  very large number of iterations $n$ in the case of deterministic hyperbolic manifolds. The case of a power-law distribution $q_m$ is however only studied analytically due to the very slow convergence of the numerical calculations in particular close to the critical point.

\begin{table*}
\bgroup
\def\arraystretch{1.5}
\begin{center}
\begin{tabular}{ccc}
$\gamma$		&   Expansion of $P_{\infty}(p_c+\Delta p)$ 
\qquad&Nature of Transition\hspace{0.5cm}		 \\
\hline
$\gamma>4$	&   $P_{\infty}(p_c)-\alpha_{\gamma}\Delta p\ln\Delta p$		& Discontinuous		\\
$\gamma=4$	&   $P_{\infty}(p_c)+\alpha_{\gamma}\Delta p\left[\ln\Delta p\right]^2$			& Discontinuous	\\
$\gamma\in(3,4)$		& $P_{\infty}(p_c)+\alpha_{\gamma}\Delta p^{\gamma-3}$ &Discontinuous \\
$\gamma=3$	& $\alpha_{\gamma}e^{-{\delta}/{\Delta p}}$	& Continuous\\
	$\gamma\in(2,3)$		&   $\alpha_{\gamma}{\Delta p}^{\hat{\beta}}$ &Continuous
\end{tabular}\\
\end{center}
\caption{Critical behavior of the  fraction of nodes $P_{\infty}(p)$ in the giant component for $p=p_c+\Delta p$ with $0<\Delta p\ll1 $, and the nature of the phase transition (continuous/discontinuous) for random hyperbolic manifolds with $q_m$ having power-law asymptotic scaling $q_m\simeq Cm^{-\gamma}$, for $m\gg 1$.}
\label{table:pinf}
\egroup
\end{table*}%

\subsection{Case of arbitrary $q_m$ distribution with  convergent second moment $\Avg{m^2}$}
\subsubsection{RG flow}
Here we consider arbitrary $q_m$ distributions with convergent second moment $\avg{m^2}$ and we derive in the continuous approximation the dependence of $\Delta T_n$ on $n$.
Our starting point is the RG  Eq.\ (\ref{RG}) that we rewrite here for convenience
\bea
T_{n+1}=F(p,T_n)=p+(1-p)\sum_{m}q_m T_n^{m-1}.
\eea
By developing this equation close to   the critical point $(p,T)=(p_c,T_c)$ and indicating with  $\Delta p=p-p_c>0$ and $\Delta T_n=T-T_c$ we get
\bea
T_{n+1}&=&F(p_c,T_c)+\left.\frac{\partial F}{\partial p}\right|_{p=p_c,T=T_c}\Delta p\nonumber \\
&&+\left.\frac{\partial F}{\partial T}\right|_{p=p_c,T=T_c}\Delta T_n+\left.\frac{\partial^2 F}{\partial p\partial T}\right|_{p=p_c,T=T_c}\Delta p \Delta T_n\nonumber \\
&&+\frac{1}{2}\left.\frac{\partial^2 F}{\partial T^2}(\Delta T_n)^2\right|_{p=p_c,T=T_c}+\ldots
\eea
with
\bea
F(p_c,T_c)&=&T_c=1,\nonumber \\
\left.\frac{\partial F}{\partial p}\right|_{p=p_c,T=T_c}&=&0,\\
\left.\frac{\partial F}{\partial T}\right|_{p=p_c,T=T_c}&=&1,\\
\left.\frac{\partial^2 F}{\partial p\partial T}\right|_{p=p_c,T=T_c}&=&-\avg{m-1},\\
\left.\frac{\partial^2 F}{\partial T^2}\right|_{p=p_c,T=T_c}&=&\frac{\avg{(m-1)(m-2)}}{\avg{m-1}}.\\
\eea
Therefore by truncating the expansion to the leading terms in $\Delta T_n$ and $\Delta p$ we can write
\bea
\Delta T_{n+1}-\Delta T_n=\hat{C}\Delta T_n \left[\Delta T_n-\hat{B}\Delta p\right]
\label{deltaT}
\eea
with  the constants $\hat{B}$ and $\hat{C}$ given by
\bea
\hat{B}=\frac{2\Avg{m-1}^2}{\avg{(m-1)(m-2)}},\\
\hat{C}=\frac{1}{2}\frac{\avg{(m-1)(m-2)}}{\avg{m-1}}.
\eea
For  $n\to \infty$ we approximate the above equation (\ref{deltaT}) in the continuous limit and we  use  $x$ to indicate the continuous approximation of $-\Delta T_n\ll 1 $, i.\,e.,  $x\simeq -\Delta T_n$. In this way we   get  the  differential equation 
\bea
\frac{dx}{dn}=-\hat{C}x[x+\hat{B} \Delta p ],
\label{xn}
\eea
with initial condition $x(0)=1-p$, whose solution is 
\bea
x(n)={\hat{B}\Delta p}\left[ \left(1+\frac{\hat{B}\Delta p}{1-p}\right)e^{\hat{C}\hat{B}(\Delta p)n }-1\right]^{-1}.
\label{xn1}
\eea

\subsubsection{Case in which both $\avg{m^2}$ and  $\avg{m^3}$ are convergent}

This case includes all distributions $q_m$ which have a convergent first, second and third moment $\avg{m},\avg{m^2}$ and $\avg{m^3}$. Consequently it includes the deterministic hyperbolic manifold with $q_{m'}=\delta_{m,m'}$ and the random hyperbolic manifold with asymptotic power-law scaling of $q_m\simeq Cm^{-\gamma}$ with $\gamma>4$.
For this case $\psi_n$ obeys the scaling relation Eq.\ (\ref{qpsin}) that we rewrite here for convenience
\bea
\psi_n=1-\hat{A}(T_c-T_n)^2=1-\hat{A} [x(n)]^2.
\eea
Therefore using Eq.\ (\ref{pinft}) we can express $P_{\infty}$ in the continuous approximation as
\bea
P_{\infty} (p)&\simeq &\exp\left[-\ln \avg{m-1} \hat{A}\int dn [x(n)]^2\right].
\eea
By inserting the expression of $x(n)$ derived in Eq.\ (\ref{xn1}) we obtain for $0<p-p_c\ll 1$
\bea
P_{\infty} (p)&\simeq
&\exp\left[-\ln(\avg{m}-1)\hat{A}_m\left(\frac{(1-p)}{\hat{C}}\right.\right.\nonumber \\
&&\left.\left.+\frac{\hat{B}\Delta p}{\hat{C}}\ln\left(\frac{\hat{B}\Delta p}{\hat{C}}\right)\right)\right]\eea
which can be also written as 
\bea
P_{\infty} (p)&\simeq &P_{\infty}(p_c)\left(\frac{\Delta p}{r}\right)^{-h \Delta p}\nonumber \\
\label{Pd1}
\eea
where $P_{\infty}(p_c), h$, and $r$ are given by  
\bea
P_{\infty}(p_c)&=&\exp\left[-\ln(\avg{m}-1)\hat{A}\frac{2}{\Avg{(m-1)(m-2)}}\right],\nonumber\\
h&=&\ln(\avg{m}-1)\hat{A}\frac{\hat{B}}{\hat{C}},\nonumber \\
r&=&\frac{\hat{C}}{\hat{B}\avg{m-1}}.
\eea
Eq.\ (\ref{Pd1}) can be further expanded for $0<\Delta p\ll1$, obtaining 
the critical behavior 
\bea
P_{\infty} (p)&\simeq &P_{\infty}(p_c)+\alpha \Delta p \left[-\ln \left(\Delta p\right)\right],
\eea
where $\alpha=P_{\infty}(p_c)h$.
This expression clearly  shows that in this case the transition is discontinuous and the fraction of nodes in the giant component has a discontinuity of  $P_{\infty}(p_c)>0$ at the critical point.
However this is not an ordinary first-order transition as ${dP_{\infty}}/{dp}$ diverges logarithmically as $\ln(\Delta p)$.\\
\begin{figure}[htbp]
\begin{center}
\includegraphics[width=0.9\columnwidth]{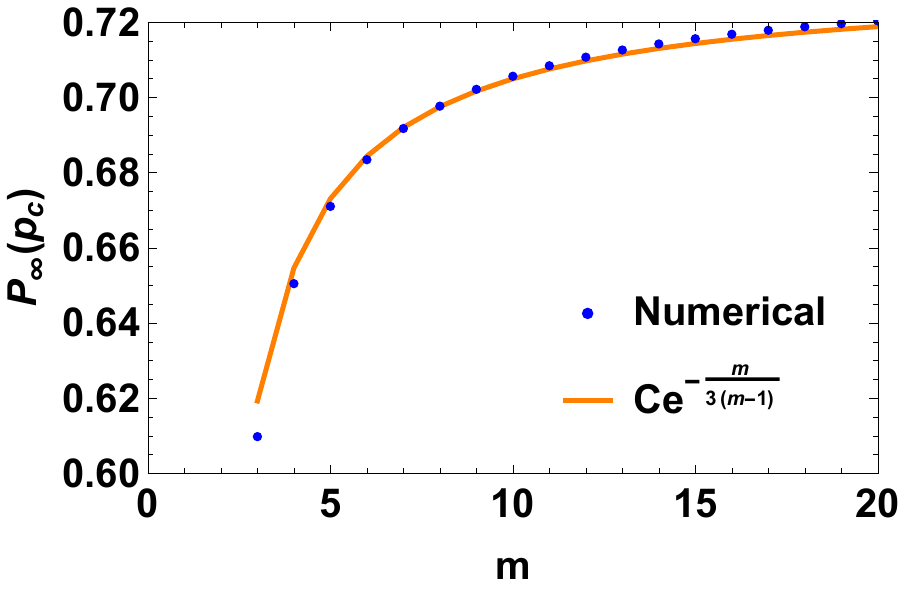}
\caption{The discontinuity $P_{\infty}(p_c)$ of the order parameter $P_{\inf}$ at the critical point $p=p_c$ is here numerically evaluated for deterministic hyperbolic manifolds with $3\leq m\leq 20$ evolved up to $n=1000$ iterations. These numerical results agree well with the analytical expression in Eq.\ (\ref{Pinfm}) up to a multiplicative constant $C=1.021\ldots$.}
\label{fig.pinf}
\end{center}
\end{figure}
\begin{figure}[htbp]
\begin{center}
\includegraphics[width=0.9\columnwidth]{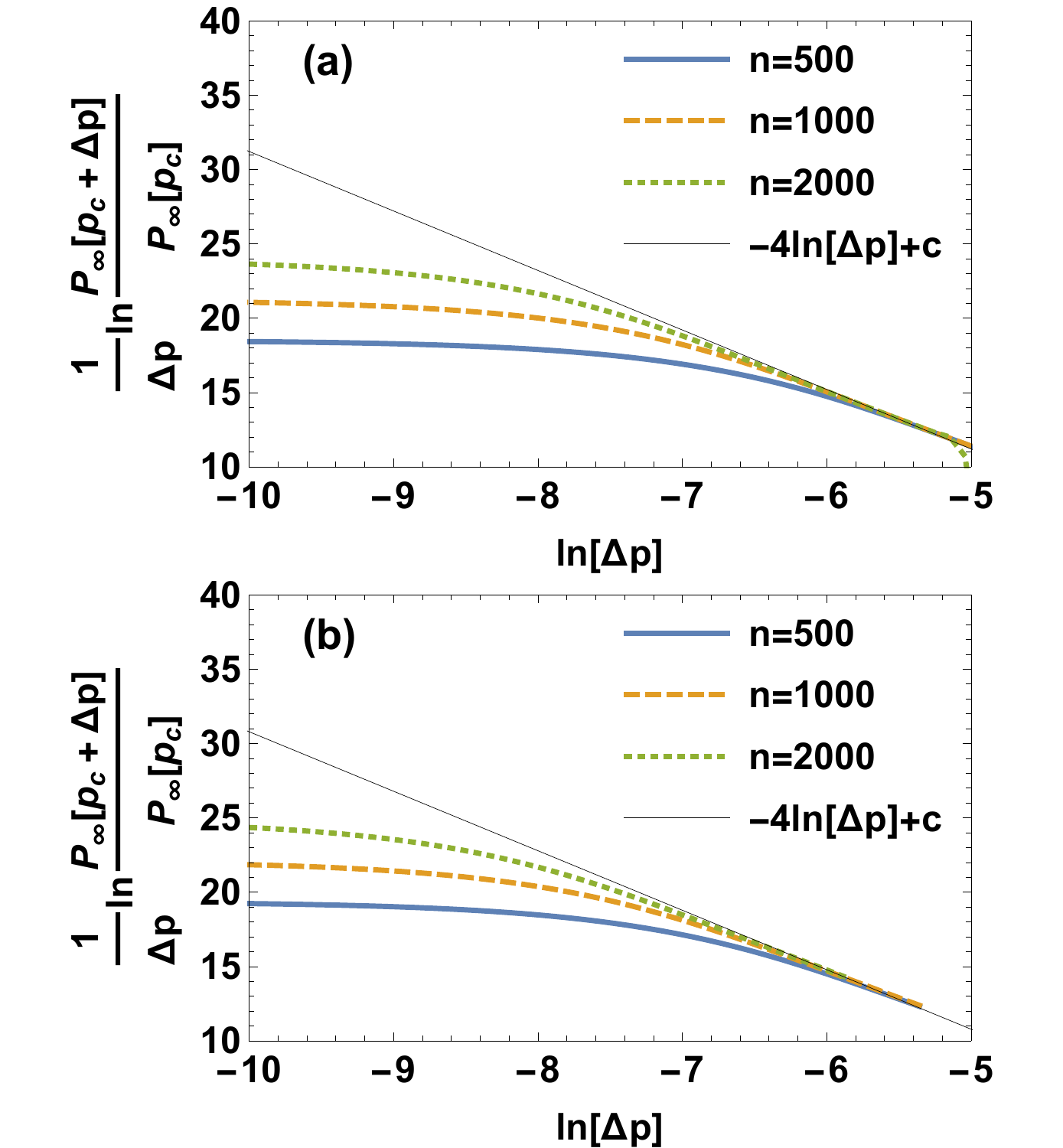}
\caption{The critical scaling of  $P_{\infty}(p)$ for $0<\Delta p=p-p_c\ll1$ is numerically investigated for deterministic hyperbolic manifolds with $m=3$ (panel a) and $m=4$ (panel b). By investigating the finite-size effects by considering a  number of iterations $n=500,1000,2000$ we validate our analytical expression Eq.\ (\ref{Pd1}). Here the constant indicates $c=4 \ln 9$ for both panels (a) and (b).}
\label{fig.crit}
\end{center}
\end{figure}
The deterministic hyperbolic manifolds having  $q_{m'}=\delta_{m',m}$ deserves some special attention.
In this case we have  that $\hat{A}$ is expressed by Eq.\ (\ref{hAm}) that we rewrite here for convenience  $$\hat{A}=\hat{A}_m=\frac{1}{6}\frac{m(m-2)}{\ln(m-1)}$$ and the size of the discontinuity $P_{\infty}(p_c)$ in the size of the giant component at the upper percolation threshold is given by 
\bea
P_{\infty}(p_c) &\simeq &\exp\left[-\frac{1}{3}\frac{m}{(m-1)}\right].
\label{Pinfm}
\eea
Moreover the constants $h$ and $r$ take the simple form 
\bea
h &=&\frac{2}{3}\frac{m(m-1)}{(m-2)},\label{h}\\
r&=&\left(\frac{m-2}{2(m-1)}\right)^2.\label{r}
\eea

These predictions can be validated by exact numerical integrations of the equations which provides the fraction $P_{\infty}$ of nodes in the giant component when the manifold includes $n$ iteration of its recursive construction.
 In Figure $\ref{fig.pinf}$ we show the numerical results obtained for the discontinuity $P_{\infty}(p_c)$ at the critical point $p=p_c$ for different values of $m$ obtained after $n=1000$ iterations and we compare these results with the expression provided in Eq.\ (\ref{Pinfm}) obtained by neglecting the non-homogeneous terms in Eq.\ (\ref{Vprime}) and performing  the continuous approximation for solving the RG flow equations.  The analytical expression differs from the exact numerical results only by a multiplicative constant $C=1.021\ldots$.
In Figure $\ref{fig.crit}$ we validate Eq.\ (\ref{Pd1}) characterizing the scaling of the order parameter $P_{\infty}(p)$ for $0<\Delta p=p-p_c\ll1$ in the case of the deterministic manifolds with $m=3,4$.
The predicted scaling is confirmed and the expression of the constant $h$ given by Eq.\ (\ref{h}) is a valid approximation, however for $m=3$ the value of the constant $r$ deviates from the predicted value given by Eq.\ (\ref{r}).

\subsubsection{Case in which $\gamma=4$ }
When the distribution $q_m$ has a power-law asymptotic scaling for $m\gg1 $ given by $q_m\simeq Cm^{-\gamma}$ and $\gamma=4$ we have shown in Sec.$\ref{sec_psi}$ that $\psi_n$ obeys the scaling
\bea
\psi_n\simeq 1-\hat{A}_{\gamma}(\Delta T_n)^{2}\ln|\Delta T_n|.
\eea
Therefore using the continuous approximation expression for $P_{\infty}(p)$ given by Eq.\ (\ref{pinft}) we obtain
\bea
P_{\infty}(p)&\simeq &\exp\left\{-\ln(\avg{m}-1) \hat{A}_{\gamma}\int_0^{\infty} dn [x(n)]^{2}\ln[x(n)]\right\}.\nonumber 
\eea
By inserting  the expression of $x(n)$ given by Eq.\ (\ref{xn1}) in the integral appearing in this exponent we obtain
\bea
&&\int_0^{\infty} dn [x(n)]^{2}\ln[x(n)]=\nonumber \\
&&= \frac{\hat{B}\Delta p}{\hat{C}}\int_{\left(1+\frac{\hat{B}\Delta p}{1-p}\right)}^{\infty} dw \frac{1}{w}\left(w-1\right)^{-2} \ln[ {\hat{B}\Delta p}(w-1)^{-1}]\nonumber \\
&&=\left[(1-p)\ln \left(\frac{1-p}{e}\right)+\frac{1}{2}\hat{B}\Delta p\left(\ln \hat{B}\Delta p\right)^2\right]
\eea
Therefore  for $0<\Delta p\ll1$ we obtain the critical behavior 
\bea
P_{\infty}(p)\simeq P_{\infty}(p_c)+\alpha_{\gamma}\Delta p\left(\ln \Delta p\right)^2
\eea
where $P_{\infty}(p_c)$ and $\alpha_{\gamma}$ are given by 
\bea
P_{\infty}(p_c)&=&\left[(\avg{m}-1)\frac{1-p}{e}\right]^{-(1-p)\hat{A}_{\gamma}},\nonumber \\
\alpha_{\gamma}&=&-\frac{\ln(\avg{m}-1)}{2}P_{\infty}(p_c)\hat{A}_{\gamma}\hat{B} .
\eea
Therefore in this case ${dP_{\infty}}/{dp}$ diverges as $(\ln \Delta p)^2$.
\subsubsection{Case in which $\gamma\in(3,4)$ }
When the distribution $q_m$ has power-law asymptotic scaling $q_m\simeq Cm^{-\gamma}$ for $m\gg1$ and $\gamma\in(3,4)$ we have shown in Sec. \ref{sec_psi} that $\psi_n$ scales like
\bea
\psi_n=1-\hat{A}_{\gamma}|\Delta T_n|^{\gamma-2}=1-\hat{A}_{\gamma} [x(n)]^{\gamma-2}
\eea
Therefore  the fraction $P_{\infty}(p)$ of nodes in the giant component can be evaluated in the continuous approximation as 
\bea
P_{\infty}(p)&\simeq &\exp\left\{-\ln (\avg{m}-1) \hat{A}_{\gamma} \int_0^{\infty} dn [x(n)]^{\gamma-2}\right\}.\nonumber \eea
The integral in the exponent of this expression can be performed by using the explicit expression of $x(n)$ given by Eq.\ (\ref{xn1}) obtaining
\bea
&&\int_0^{\infty} dn [x(n)]^{\gamma-2}=\nonumber \\
& =&\int_0^{\infty} dn {(\hat{B}\Delta p)^{\gamma-2}}\left[ \left(1+\frac{B\Delta p}{1-p}\right)e^{\hat{B}\hat{C}\Delta p n}-1\right]^{-(\gamma-2)}\nonumber \\
& =& \frac{(\hat{B}\Delta p)^{\gamma-3}}{\hat{C}}\int_{\left(1+\frac{B\Delta p}{1-p}\right)}^{\infty} dw \frac{1}{w}\left[ w-1\right]^{-(\gamma-2)}\nonumber \\
& \simeq & \frac{(1-p)^{\gamma-3}}{\hat{C}(\gamma-3)}-D_{\gamma} (\Delta p)^{\gamma-3},
\eea
where $D_{\gamma}$ is a constant.
Therefore  for $0<\Delta p\ll1$ we obtain the critical behavior 
\bea
P_{\infty}(p)\simeq P_{\infty}(p_c)+\alpha_{\gamma}(\Delta p)^{\gamma-3}
\eea
characteristic of {\em hybrid phase transitions} where $P_{\infty}(p_c)$ and $\alpha_{\gamma}$ are given by 
\bea
P_{\infty}(p_c)&=&\exp\left[-\ln(\avg{m}-1)\hat{A}_{\gamma}\frac{(1-p)^{\gamma-3}}{\hat{C}(\gamma-3)}\right],\nonumber \\
\alpha_{\gamma}&=&P_{\infty}(p_c)\ln(\avg{m}-1)\hat{A}_{\gamma}D_{\gamma}
\eea
Therefore in this case a discontinuous phase transition is expected. However  ${dP_{\infty}}/{dp}$ diverges as $(\Delta p)^{\gamma-4}$. 
\subsection{Convergent $\Avg{m}$, divergent $\Avg{m^2}$ }
\subsubsection{Case in which $\gamma=3$ }
Here we consider random hyperbolic manifolds with  a $q_m$ distribution having an asymptotic power-law scaling $q_m\simeq Cm^{-\gamma}$ for $m\gg1$ with $\gamma=3$.
In this case, by expanding the RG Eq.\ (\ref{RG}) close to the upper percolation threshold for small $\Delta T_n$ and $\Delta p$  we obtain
\bea
\Delta T_{n+1}-\Delta T_n=\Delta T_n \left[-\Avg{m-1} \Delta p+d_{\gamma}(\Delta T_n)\ln |\Delta T_n| \right],\nonumber
\eea
where we have truncated the expansion by taking only the leading terms in $\Delta p$ and $\Delta T_n$. 
In the continuous approximation we get the differential equation 
\bea
\frac{dx}{dn}=-\bar{C}x[x\ln x+\bar{B} \Delta p ]
\label{xn2}
\eea
with initial condition $x(0)=1-p$.
This differential equation does not have an explicit analytical solution. In order to find an  approximate solution we consider two different ranges of values of $n$.  The first range of values of $n$ is  $n<n^{\star}$ where $n^{\star}$ satisfies $x(n^{\star})\ln x(n^{\star})=B\Delta p$. In this regime we observe $x(n)\ln x(n)>B\Delta p$.  The second range is $n>n^{\star}$ where we observe $x(n)\ln x(n)<B\Delta p$.
Therefore we can integrate Eq.\ (\ref{xn2}) from $x(0)$ to $x(n^{\star})$ and consider the first term of the expansion for $x(n)\ln x(n)/(B\Delta p)\ll 1$. Subsequently we can integrate Eq.\ (\ref{xn2}) from  $x(n^{\star})$ to a generic $x(n)<x(n^{\star})$ and consider only the first term of the expansion for  $B\Delta p/[x(n)\ln x(n)]\ll 1$. 
We note that $n^{\star}$ is determined by  
\bea
x(n^{\star})=\frac{B\Delta p}{\ln[B\Delta p]}.
\eea
For $n<n^{\star}$ we obtain that the leading term of the solution of Eq.\ (\ref{xn2}) reads
\bea
\mathrm{Ei}(\ln(x(n)))-\mathrm{Ei}(\ln (1-p))=-\bar{C}n,
\eea
where Ei$(z)$ indicates the exponential integral function. We note  that as expected this solution obtained by considering the leading term in $x(n)\ln(x)/[B\Delta p]$ is independent of $\Delta p$ as expected.
For $n>n^{\star}$ we obtain instead 
\bea
x(n)=x(n^{\star})e^{-\bar{B}\bar{C}\Delta p(n-n^{\star})}.\label{xn3}
\eea
As we have shown in Sec. $\ref{sec_psi}$ the asymptotic scaling of $\psi_n$ for $\gamma=3$ is given by 
\bea
\psi_n\simeq 1-\hat{A}_{\gamma} x(n) \ln x(n).
\eea
Therefore, by inserting this scaling in Eq.\ (\ref{pinft}) we get
\bea
P_{\infty}&\simeq &\exp\left\{-\ln (\avg{m}-1) \hat{A}_{\gamma} \int_0^{\infty} dn x(n)\ln x(n)\right\}\nonumber \\
&&\exp\left\{\phi-\ln (\avg{m}-1) \hat{A}_{\gamma}\int_{n^{\star}}^{\infty} dn x(n)\ln{ x(n)}\right\}\nonumber
\label{P4_ns}
\eea
where in the second expression the constant $\phi$ is given by  
\bea
\phi=\int_{0}^{n^{\star}} dn x(n)\ln x(n).
\eea
By using Eq.\ (\ref{xn3}) to express $x(n)$ for $n>n^{\star}$ in the equation for $P_{\infty}(p)$ we can determine the critical scaling of $P_{\infty}(p)$ close to the upper percolation threshold, given by  
\bea
P_{\infty}(p)\simeq \alpha_{\gamma}e^{-{\delta}/{\Delta p}}
\eea
where $\delta={x(n^{\star})}/({\bar{B}\bar{C}})$  and $\alpha_{\gamma}=e^{\phi}$.
Therefore in this case the transition is continuous and displays a critical behavior with an effective {\em `infinite dynamical exponent'} which has been observed also for percolation in random scale-free networks \cite{Doro_crit}.

\subsubsection{Case $\gamma\in (2,3)$}
Here we consider random hyperbolic manifolds with  a $q_m$ distribution having an asymptotic power-law scaling $q_m\simeq Cm^{-\gamma}$ for $m\gg1 $ with $\gamma\in (2,3)$.
In this case, proceeding as in the previous cases and  expanding the RG Eq.\ (\ref{RG}) close to the upper percolation threshold for small $\Delta T_n$ and $\Delta p$  we obtain
\bea
\Delta T_{n+1}-\Delta T_n=\Delta T_n \left[-\Avg{m-1} \Delta p+d_{\gamma}(\Delta T_n)^{\gamma-2}\right].\nonumber
\eea
In the continuous approximation we get the differential equation 
\bea
\frac{dx}{dn}=-d_{\gamma}x[x^{\gamma-2}+\tilde{B} \Delta p],
\label{xn3}
\eea
with initial condition $x(0)=1-p$ and with $\tilde{B}$ given by 
\bea
\tilde{B}=\frac{\Avg{m}-1}{d_{\gamma}}.
\eea
Equation $(\ref{xn3})$ has solution
\bea
x(n)=\left[\left ((1-p)^{2-\gamma}+\frac{1}{\tilde{B}\Delta p}\right)e^{d_{\gamma}\tilde{B}(\Delta p)n}-\frac{1}{\tilde{B}\Delta p}\right]^{-1/(\gamma-2)}.\nonumber
\label{xnl}
\eea
Since, as we have shown in Sec. $\ref{sec_psi}$ for this range of $\gamma$ values, $\psi_n$ obeys the scaling
\bea
\psi_n\simeq1-\hat{A}_{\gamma} [x(n)]^{\gamma-2},
\eea
by using Eq.\ $(\ref{pinft})$ we can express the fraction $P_{\infty}$ of nodes in the giant component as 

\bea
\ln P_{\infty}(p)&=&\exp\left\{-\hat{A}_{\gamma}\ln (\avg{m}-1)\int_{0}^{\infty}dn  [x(n)]^{\gamma-2}\right\}\nonumber
\eea
Finally by inserting the expression for $x(n)$ given by Eq.\ (\ref{xnl}) we get for $0<\Delta p\ll 1$
\bea
&&\int_{0}^{\infty}dn  [x(n)]^{\gamma-2}=\nonumber \\
&&\int_{0}^{\infty}dn \left[\left ((1-p)^{2-\gamma}+\frac{1}{\tilde{B}\Delta p}\right)e^{d_{\gamma}\tilde{B} (\Delta p)n}-\frac{1}{\tilde{B}\Delta p}\right]^{-1}\nonumber \\
&=&-\frac{\ln\left[\hat{B}\Delta p(1-p)^{2-\gamma}\right]}{d_{\gamma}\left[1+\tilde{B}\Delta p (1-p)^{2-\gamma}\right]}\nonumber \\
&&\simeq-\frac{1}{d_{\gamma}}\ln \Delta p+D_{\gamma}.
\eea
where $D_{\gamma}$ is a constant.
Therefore the transition is continuous with
 the critical behavior 
\bea
P_{\infty}(p)\simeq \alpha_{\gamma}(\Delta p)^{\hat{\beta}}
\eea with $\alpha_{\gamma}=\exp[-\hat{A}_{\gamma}\ln(\avg{m}-1)D_{\gamma}]$ and the dynamical exponent
$\hat{\beta}$ given by 
\bea
\hat{\beta}=\frac{\hat{A}_{\gamma} \ln \avg{m-1}}{d_{\gamma}}.
\eea
\section{Conclusions}
In conclusion, we have presented a comprehensive renormalization group study of link percolation in two-dimensional hyperbolic manifolds. 
The considered manifolds are the skeletons of two-dimensional cell complexes and are iteratively constructed by gluing polygons to links.
In particular we have considered the deterministic hyperbolic manifolds formed by identical polygons of size $m$ and random hyperbolic manifolds constructed by gluing polygons of size $m$ drawn from a distribution $q_m$.
Link percolation on deterministic hyperbolic manifolds with $m=3$ (Farey graphs) has been previously shown \cite{hyperbolic_Ziff} to display a discontinuous phase transition at the upper percolation threshold.
Here we extend these results by predicting that for any fixed value of $m$ the transition remains discontinuous and we analytically estimate the discontinuity, finding very good agreement with exact numerical results.
For deterministic hyperbolic manifolds, the critical behavior deviates from the one expected for a first-order phase transition as the derivative of the order parameter diverges logarithmically close to the upper percolation threshold $p_c$.
The study of random hyperbolic manifolds with power-law distribution $q_m$ shows  a rich set of different  universality classes depending on the value of $\gamma$. In particular for $\gamma\in (3,4)$ we predict a hybrid phase transition while for $\gamma \in (2,3]$ the transition is predicted to be continuous.
Therefore this work reveals that  in hyperbolic manifolds a power-law $q_m$ distribution can have profound effect on the universality class of the link percolation transition at the upper percolation threshold. In this work this phenomena is compared with the well-known effect that power-law degree distributions have on the percolation properties of random networks. This comparison allows us to propose a mathematical mapping between the equation for the percolation probability and the equation determining the probability that by following a link we reach a node in the giant component of a random network. However the equations determining the order parameter in the hyperbolic manifold do not appear to have an equivalent counterpart in the percolation of random networks.

We hope that this work will stimulate further research in the interplay between network geometry and dynamics and in particular in the properties of percolation in hyperbolic and non-amenable networks.

\bibliographystyle{apsrev4-1}
\bibliography{biblio_cell2d}
\end{document}